\documentclass{article}
\usepackage{geometry}
\usepackage{xcolor}
\usepackage{amsmath}
\usepackage{amssymb}
\usepackage{bbm}
\usepackage{graphicx} 
\usepackage{cprotect}
\usepackage{makecell}
\usepackage{tabularx}
\usepackage{multirow}
\usepackage{hyperref}
\usepackage{prettyref}
\usepackage{booktabs}

\usepackage{colortbl}
\usepackage{lineno}
\usepackage{authblk}
\usepackage{placeins}
\usepackage[numbers]{natbib}
\usepackage{fontspec}  
\usepackage[font=footnotesize ,labelfont=bf]{caption}

\DeclareMathOperator*{\argmin}{arg\,min}

\newrefformat{fig}{Figure~\ref{#1}}
\newrefformat{tab}{Table~\ref{#1}}
\newrefformat{sec}{Section~\ref{#1}}
\newrefformat{eq}{Equation~\ref{#1}}

\usepackage{changes}
\definecolor{YB}{RGB}{0,0,200}
\definecolor{BZ}{RGB}{0,200,0}
\definecolor{RD}{RGB}{200,0,0}
\definechangesauthor[color=YB]{YB}
\definechangesauthor[color=BZ]{BZ}
\definechangesauthor[color=RD]{RD}

\newcommand{\ssbquote}[1]{``#1''}

\newcommand{\ms}{\,\textnormal{ms}}

\newrefformat{tab}{Table~\ref{#1}}

\title{\textsc{Sym\textbf{Seq}Bench}: a unified framework for the generation and analysis of rule-based symbolic sequences and datasets}
\author[1]{\textbf{Zajzon, Barna}\thanks{b.zajzon@fz-juelich.de}}
\author[2]{\textbf{Bouhadjar, Younes}}
\author[2, 3]{\textbf{Fabre, Maxime}}
\author[2, 4]{\textbf{Schmidt, Felix}}
\author[1, 4]{\textbf{Ostendorf, Noah}}
\author[2, 6]{\textbf{Neftci, Emre}}
\author[1, 5]{\textbf{Morrison, Abigail}}
\author[7, 8]{\textbf{Duarte, Renato}}
\affil[1]{Institute for Advanced Simulation (IAS-6), Jülich Research Centre, Germany}
\affil[2]{Peter-Grunberg Institute, Neuromorphic Software Ecosystems (PGI-15), Jülich Research Centre, Germany}
\affil[3]{Groningen Cognitive Systems and Materials Center (CogniGron), University of Groningen}
\affil[4]{RWTH Aachen University, Aachen, Germany}
\affil[5]{Department of Computer Science 3 - Software Engineering, RWTH Aachen University, Aachen, Germany}
\affil[6]{Department of Electrical Engineering and Information Technology, RWTH Aachen University, Aachen, Germany}
\affil[7]{Center for Neuroscience and Cell Biology (CNC-UC), University of Coimbra, Portugal}
\affil[8]{Center for Innovative Biomedicine and Biotechnology (CIBB-UC), University of Coimbra, Portugal}

\begin{document}

\maketitle

\begin{abstract}
Sequential structure is a key feature of multiple domains of natural cognition and behavior, such as language, movement and decision-making. Likewise, it is also a central property of tasks to which we would like to apply artificial intelligence. It is therefore of great importance to develop frameworks that allow us to evaluate sequence learning and processing in a domain agnostic fashion, whilst simultaneously providing a link to formal theories of computation and computability.   
To address this need, we introduce two complementary software tools: \verb|SymSeq|, designed to rigorously generate and analyze structured symbolic sequences, and \verb|SeqBench|, a comprehensive benchmark suite of rule-based sequence processing tasks to evaluate the performance of artificial learning systems in cognitively relevant domains. 
In combination, \verb|SymSeqBench| offers versatility in investigating sequential structure across diverse knowledge domains, including experimental psycholinguistics, cognitive psychology, behavioral analysis, neuromorphic computing and artificial intelligence. 
Due to its basis in Formal Language Theory (FLT), \verb|SymSeqBench| provides researchers in multiple domains with a convenient and practical way to apply the concepts of FLT to conceptualize and standardize their experiments, thus advancing our understanding of cognition and behavior through shared computational frameworks and formalisms. The tool is modular, openly available and accessible to the research community.

\end{abstract}

\section{Introduction}

Symbols and symbol systems are central constructs in both cognitive sciences and theoretical computer science, providing a unique formal link between cognition and computation \citep{stephen_jose__hanson_ea1b50cb}. This relation, manifested as a branch of the mathematical theory of computation known as Formal Language Theory (FLT), grounds the investigation of complex cognitive, psychological and behavioral processes onto a systematic terminology and set of conventions for describing generative rule systems and the structures they generate \citep{gerhard_ja_ger_bbb90244, w_tecumseh_fitch_ece3a959}. From simple pattern perception to language and sequential decision-making, such formal systems allow the algorithmic specification of complex cognition and can be used to understand, model and analyze cognition and behavior \citep{roma_patel_9b004cc0, christopher_i__petkov_3a05f34b, steven_t__piantadosi_50d37955}.

Evaluating sequence learning and processing has long been central to advancing our understanding of both artificial and natural intelligence, from Lashley's seminal work on serial order and the syntax of action \citep{lashley_problem_1951} to contemporary large-scale neural sequence models. Accounting for the full scope of sequence processing is nevertheless challenging: system-specific constraints and the many dimensions along which task characteristics and model demands vary mean that most tasks probe only a limited portion of the problem space. Canonical experimental paradigms and principled benchmarks make it possible to address this challenge by providing shared tasks and evaluation protocols that allow systematic comparison and reveal key limitations in learning and generalization \citep{hurlstone_memory_2014,sculley2018winners,Geirhos2020,Raji2021}. The rise of linguistically-competent and computationally proficient artificial intelligence \citep{nova_spivack_8421a206} has led to renewed interest in tasks and metrics that can evaluate systems' capacity to learn and generalize complex structured sequences \citep{aarohi_srivastava_7ed2343a, james_fodor_5915e049, ankur_sikarwar_ace45230}. Ideally, and given that human intelligence remains the ultimate reference point, such tasks should be grounded in cognitive theory and capture relevant aspects of human reasoning, behavior and cognition \citep{brenden_m__lake_2ea4879c, jos__hern_ndez_orallo_93f4b461, kriegeskorteCognitiveComputationalNeuroscience2018, palmeriModelbasedCognitiveNeuroscience2017}. We argue that moving beyond monolithic approaches or discrete language-theoretic classes toward a more fluid understanding of temporal processing requires tools that capture its multifaceted nature and complexity at multiple scales, enabling the principled design of tasks with controllable sequence complexity. These allow researchers to scrutinize the properties of sequential structure, generative capacity and the requirements it imposes on the computational machinery.

Strongly inspired by decades of human psycholinguistic research, the tools we present here address these needs by providing a comprehensive framework for the specification, generation, manipulation, and analysis of symbolic sequences and corresponding embeddings, as well as a benchmark suite of datasets and tasks that can be used to evaluate the performance of artificial cognitive systems. With applications spanning experimental psychology, behavioral analysis, neuromorphic computing, and artificial intelligence, among others, our framework aims to facilitate interdisciplinary research and the adoption of formal constructs to multiple scientific domains that investigate structured sequential data \citep{daniel_silver_e3edf12e, gerard_o_regan_0ab1b793}. We illustrate the scope and applicability with concrete use-cases in areas such as dataset generation for artificial grammar learning experiments, the analysis and comparison of behavioral ethograms in different experimental conditions, the evaluation of neuromorphic, artificial neural network architectures, and the mechanistic dissection of neural sequence models.

While other tools and approaches have been proposed over the years that address specific aspects of symbolic sequence processing, cognitive modeling and computational benchmarking, they largely do so with a narrow focus. For example, specialized software exists to automatically generate and select string sets for artificial grammar learning (AGL) experiments  \citep{Bailey2008}, to evaluate large language models’ meta-learning and generalization from minimal exposure  \citep{ekin_aky_rek_b020c7d7}, to systematically deploy standardized neuroscience experiments and tasks \citep{manuel_molano_maz_n_c01c84dc} or to investigate neuromorphic embodied agents’ performance in standardized environments and tasks \citep{philipp_weidel_a8eab012, jakob_jordan_07d4b27f}, among others. An integrated, broadly applicable tool requires a higher level of abstraction and generalization. The adoption of FLT concepts, as we propose here, has proven insightful for predicting the generalization limits of artificial neural networks \citep{Deletang22_chomsky}, proposing architectural augmentations that could increase their expressiveness \citep{Deletang22_chomsky, Sima2020} or to expose computational similarities and differences between biological and artificial intelligence \citep{Fitch2014}. 

These studies provide powerful tools and valuable theoretical insights into the algorithmic and computational properties and limitations of neural architectures designed to cope with sequential structural regularities. They demonstrate the importance of appropriate computational formalisms and benchmark tasks, but are not easily generalizable across research domains. The ubiquity of temporal structure in animal cognition and behavior, as well as the power of adopting formal, systematic constraints and notations across scientific domains, urges a more substantive integration and the development of tools that can homogenize experimental paradigms and metrics, providing a shared conceptual framework to interpret and analyze different aspects of biological and artificial intelligence.

Unlike existing specialized software, \verb|SymSeqBench| offers a comprehensive, multi-disciplinary approach to analyze human, animal, and artificial intelligence, from designing psycholinguistic experiments and evaluating neuromorphic architectures to dissecting the syntactic structure of behavioral and biological sequences. Below, we explain the theoretical bases and formalisms linking the approaches proposed throughout the paper, we proceed to explain how the tool is organized and implemented, followed by a set of concrete use-cases demonstrating its scope and applicability across a wide range of problem domains.

\subsection{Theoretical Foundations}
\label{sec:foundations}
We consider sequences as elements of (potentially infinite) formal languages, defined over a finite alphabet and generated by a formal grammar. A formal \textbf{language} $\mathcal{L}$ comprises the set of all \textbf{words} or \textbf{strings} $\mathcal{S}$ that can be formed by following a generative \textbf{grammar} $\mathcal{G}$, i.e. a set of production rules. Each string $S_{i}$ comprises a finite sequence of \textbf{symbols} $\sigma_{i}$, drawn from a finite \textbf{alphabet} $\mathcal{A}$. The generative grammar $\mathcal{G}$ defines the \textbf{syntax} of the language, specifying how symbols can be combined to form valid strings and string-sets, and can be described using various formalisms, such as regular expressions, context-free grammars, or more expressive representations like context-sensitive or unrestricted grammars. 

The notion of syntax is the defining anchor, as it captures, quantifies and formalizes all aspects pertaining to temporal structure: the systematic patterns governing how elements are ordered and combined over time. By formalizing syntax through generative grammars, we provide a principled framework for specifying, manipulating, and analyzing the temporal dependencies that characterize complex sequences. 

Due to the deep relationship between formal languages and abstract automata  \citep{noam_chomsky_f23c6dd3, noam_chomsky_89248260}, a grammar can be represented by a corresponding automaton $\mathcal{A}$, such that $\mathcal{L}(\mathcal{G})=\mathcal{L}(\mathcal{A})$, where $\mathcal{L}$ denotes the language generated or recognized. For every grammar, there exists a corresponding automaton that recognizes or generates the same language. As such, to simplify the terminology, we often describe generative grammars using the tuple notation of their corresponding automata. We will primarily be dealing with regular grammars, which can be represented as $\mathcal{G}=<\mathcal{Q}, \mathcal{A}, \mathcal{T}, q_{0}, \mathcal{F}>$, where $\mathcal{Q}$ is a finite set of states, $\mathcal{A}$ is the alphabet, $\mathcal{T}$ is the state transition table, $q_{0}$ and $\mathcal{F} \subseteq \mathcal{Q}$ are the subsets of start and terminal states, respectively. Additionally, for convenience and without loss of generality, we typically consider grammatical structures as directed graphs, where nodes represent states, edges represent transitions between states, and edge labels represent symbols from the alphabet. We also use this notation interchangeably with a Markov chain formulation where symbols label states rather than transitions; the correspondence between these two equivalent representations is detailed in \prettyref{sec:symseq-main}.

The universality of symbolic sequences and the underlying theoretical notions extends across diverse scientific domains, reflecting the fundamental role of structured temporal patterns in both natural and engineered systems. In molecular biology, stochastic context-free grammars serve as standard tools for RNA secondary structure prediction and analysis \citep{10.1371/journal.pcbi.1009291}. In neuroscience, sequential patterns of neural activity or behavioral states can be characterized using formal language frameworks \citep{w_tecumseh_fitch_ece3a959, chiTemplateBasedSpikePattern2007, brinkmanMetastableDynamicsNeural2022}. In cognitive psychology, hierarchical phrase structure grammars capture the compositional nature of language and action planning \citep{Makuuchi2012, ChristiansenChater:2015, jackendoffFoundationsLanguageBrain2004}. This domain-general applicability stems from the abstract nature of formal grammars: by imposing no constraints on the semantic interpretation of symbols, which can represent any abstract state or event, the framework applies to any sequentially structured phenomenon.

We can thus define different levels of generation and analysis where we can either quantify or manipulate the qualities and characteristics of syntactic structure: (i) individual symbols or tokens $\sigma_{i} \in \mathcal{A}$; (ii) individual sub-sequences or strings $S_{i} \in \mathcal{S}$; (iii) collections of valid strings or string-sets $[S_{i}, …, S_{T}] \in \mathcal{S}$; (iv) generative grammars $\mathcal{G}_{x}$; and complete (potentially infinite) languages $\mathcal{L}$. The implementations we propose provide the ability to manipulate and/or analyse the properties at each of these levels independently and, by imposing no constraints or specific meanings to the nature of the symbols (which can represent any abstract state), are applicable to a very broad range of sequentially structured data. 

It is important to distinguish our use of certain formal language theoretic concepts from their traditional interpretations, particularly the notion of ambiguity. In formal language theory, ambiguity typically refers to the existence of multiple parse trees for a single string under a given context-free grammar and thus refer to a structural property of the grammar itself. In our framework, we use the term more loosely to capture uncertainty or variability at multiple levels: symbol-level ambiguity (e.g., noisy or partially observable tokens), string-level ambiguity (e.g., multiple valid interpretations of the same surface form), and grammar-level ambiguity (e.g., underspecification in the generative rules). This generalized notion of ambiguity is essential for modeling realistic sequence processing scenarios where uncertainty pervades observation, representation, and inference.

In the following sections, we describe the implementation of this multi-level framework and demonstrate how it enables systematic exploration of sequence complexity across scales, providing both theoretical insights and practical tools for designing cognitively grounded experimental paradigms.

\section{Architecture, design and implementation}
\verb|SymSeqBench| is an open-source, modular Python framework for symbolic sequence processing, integrating grammar-based sequence generation with synthetic and real-world datasets, along with a wide range of structural and statistical analysis within a unified system. The framework divides responsibilities across two specialized components: \verb|SymSeq|\footnote{\url{https://github.com/symseqbench/symseq}} for defining, generating and analyzing symbolic sequences and tasks, encapsulated into a standardized interface, and \verb|SeqBench|\footnote{\url{https://github.com/symseqbench/seqbench}} for transforming and embedding (grounding) symbolic representations and assigning a functional meaning to symbols. This layered organization supports both synthetic and user-provided data, enabling analyses that range from abstract structural properties to distributed vector representations compatible with machine learning and biological models of computation. The advantage of this separation is that researchers interested primarily in symbolic generation and analysis can use the lightweight \verb|SymSeq| component without the additional overhead and dependencies of \verb|SeqBench|. At the same time, the flexible and user-friendly architecture emphasizes reproducibility and customization, while optimizations and a backend-agnostic design ensure scalability across different experimental and computational settings. 

\begin{figure}[!ht]
    \centering
    \makebox[\textwidth][c]{
        \includegraphics[width=180mm]{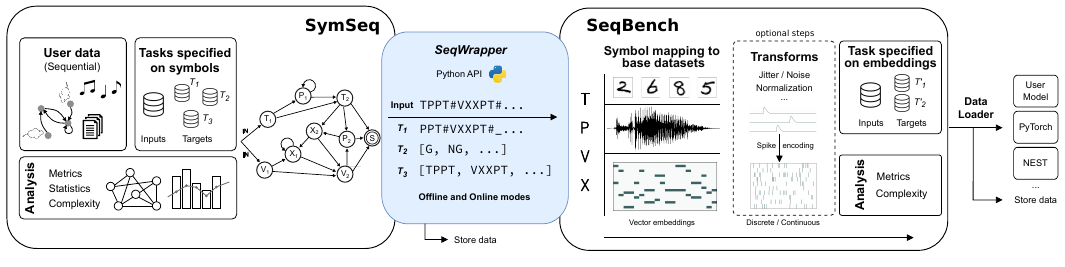}
    }
    \cprotect\caption{\textbf{Conceptual overview of \texttt{SymSeqBench}, the individual components and their interactions.} \verb|SymSeq| focuses on handling symbolic sequences, specifying processing tasks and enabling the structural analysis of both synthetic and user-provided sequences. SeqWrapper conveniently bundles \verb|SymSeq| outputs --- such as input/output sequences and generator objects --- and represents the primary interface for downstream processing. From such wrapper objects, \verb|SeqBench| provides functionality for grounding symbolic representations, assigning each symbol to a corresponding discrete- or continuous-time distributed representation, along with additional tasks and analysis that are based on embedded representations. By combining a tool-agnostic design with multiple backends, \verb|SeqBench| enables fast and efficient integration with simulation frameworks such as PyTorch and NEST. }
\end{figure}

\subsection{\texttt{SymSeq}: Symbolic sequence generation and analysis}


\verb|SymSeq| provides the foundation for defining, generating, and analyzing symbolic sequences, structuring them into relevant computational tasks and specifying all the relevant mappings. It contains core data structures for grammar and sequence generation, a diverse library of artificial language generators, interfaces for parsing user-supplied data, and a comprehensive suite of analysis metrics spanning multiple structural and linguistic scales. In addition, \verb|SymSeq| formalizes task targets in terms of language recognition and language transduction, thereby supporting both \textit{abstract rule generalization} and \textit{structure-sensitive transformations}. Together, these elements establish a flexible and extensible framework for modeling and evaluating symbolic sequence processing across domains.

Leveraging a modular design, \verb|SymSeq| allows users to use individual components and functions in isolation or instantiate complete input/output datasets from configuration files, thereby significantly reducing time-to-data in most practical scenarios. These include small, carefully hand-crafted datasets commonly used in AGL experiments (\prettyref{sec:apps-agl}), as well as large-scale datasets for neuromorphic and AI applications (\prettyref{sec:apps-benchmarks}), which can be pre-generated efficiently through parallelized routines or prepared for online generation at a later stage. In both cases, the data is exposed through a simple wrapper interface that enables flexible (custom, user-defined) downstream processing and seamless integration with \verb|SeqBench|.

Although certain features of \verb|SymSeq| are shared with other libraries and codebases accompanying related publications, these alternatives are typically not actively maintained, depend on proprietary software, or are too narrowly tailored to specific contexts to be broadly deployed. For instance, the Matlab library AGL StimSelect \citep{Bailey2008} for the generation of controlled AGL experimental datasets is not actively developed while the more recent AGSuite \citep{cook_agsuite_2017} is entirely browser-based, limited to small datasets and does not include grammar-generation functionality. On the other hand, works on formal languages in artificial neural networks usually provide custom implementation of their tasks based on the study's specific requirements \citep{Deletang22_chomsky} and are not easily generalizable beyond their intended application domain. 

While tools like NeuroGym \citep{manuel_molano_maz_n_c01c84dc} provide flexible and extensible reinforcement learning environments for cognitive neuroscience with predefined decision-making tasks, they lack the flexible symbolic framework and controllable complexity generation that \verb|SymSeq| offers for sequence processing. Similarly, experimental software such as PsychoPy \citep{Peirce2019} and PsyToolkit \citep{stoet_psytoolkit_2010,stoet_psytoolkit_2017} excel at stimulus presentation and behavioral data collection in human experiments but do not provide grammar-based task generation, sequence complexity control, or the multi-scale analysis metrics central to \verb|SymSeq|'s design. 
To the best of our knowledge there is no related software that matches the scope that abstract symbolic representations and FLT endow or the wide range of analysis tools included in \verb|SymSeq|. In the following, we describe the main components of the tool.

\begin{figure}[!ht]
    \centering  
    \makebox[\textwidth][c]{
        \includegraphics[width=180mm]{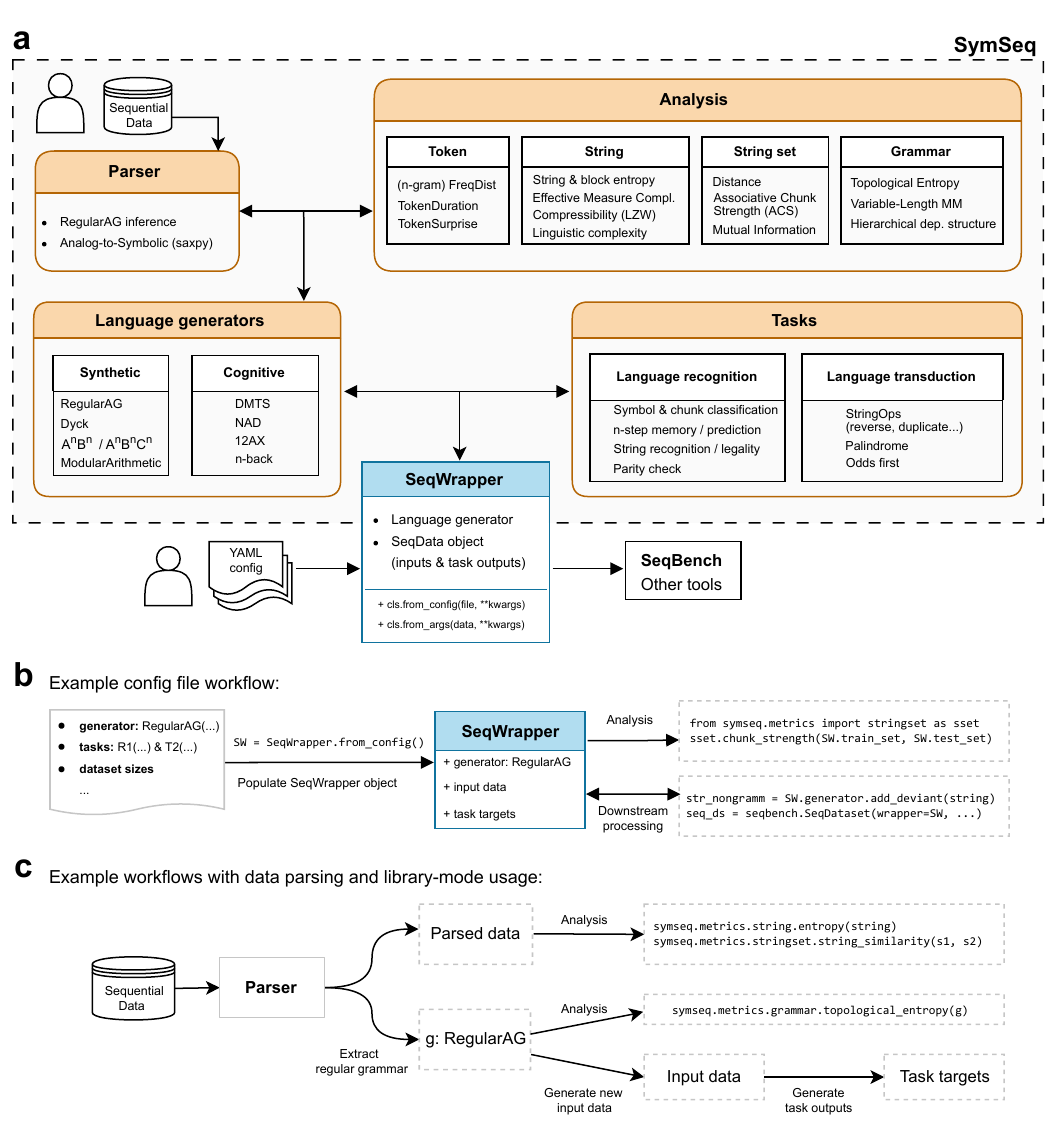}
    }
    \cprotect\caption{\textbf{Schematic overview of \texttt{SymSeq} and typical use-case workflows.}
    \textbf{a)} The library consists of four conceptual components that interact through standard Python data structures representing symbolic sequences: The \textit{Parser} accepts sequential data from the user, either as continuous time-series from which symbolic sequences are built, or as discrete symbolic sequences (string-sets). From first-order transition probabilities, the parser extracts a regular (finite-state) grammar and, resorting to the large battery of \textit{analysis} metrics, provides extensive and detailed information about the complexity of the sequences and the underlying generative grammar. \textit{Language generators} comprise various classes for creating task-specific sequences, grouped into synthetic and cognitive/behavioral categories. Both the generated strings and their underlying structure (e.g., transition table) can be directly used by functions in \textit{Analysis} which provide complexity metrics at multiple scales.
    The \textit{Tasks} module is a collection of functions to create target labels for the provided input sequences, organized into recognition and transduction categories; Finally, \textit{SeqWrapper} serves as the unified input-output interface of \verb|SymSeq|, coordinating flow logic and instantiating a complete and ready-to-use experimental setup and dataset.
    \textbf{b)}
    SeqWrapper objects can be instantiated from YAML configuration files, allowing subsequent analysis or manipulation of datasets, as well as downstream processing using \verb|SeqBench|.
    \textbf{c)}
    User-provided sequential data in symbolic form can be analyzed directly or used to infer a regular grammar and create input/output datasets from the corresponding language generator.
    }
\end{figure}

\subsubsection{Main components}\label{sec:symseq-main}

\paragraph[short]{\textbf{Parser}} 
Reads user-provided sequential data either in a symbolic or continuous time-series form. Symbolic data can be directly analyzed or used to infer regular grammars and instantiate a corresponding language generator object (see also \prettyref{sec:behavior-syntax}). Continuous data is first converted to symbolic representation using the saxpy library \citep{senin2018grammarviz}, which is based on the SAX algorithm \citep{lin_finding_nodate} to discretize the signal and transform it to a sequence of strings based on a predefined alphabet. The resulting data can be further processed similarly to any internal symbolic sequence. After inferring the underlying structure and mapping it to a language generator (e.g., regular grammar), newly generated strings can be converted back to analog signals using \verb|SeqBench|. Thus, \verb|SymSeqBench| provides a full  \verb|continuous| $\rightarrow$ \verb|symbolic|  $\rightarrow$ \verb|grammar|  $\rightarrow$ \verb|synthetic continuous| pipeline for generating synthetic time series with prescribed and controllable complexity. Note that grammar induction at this stage is limited to regular, finite-state structures (see \prettyref{sec:discussion} for a more extended explanation).

\paragraph[short]{\textbf{Language generators}}
This central component generates complete sets of structured symbolic sequences in the form of artificial languages (see \prettyref{sec:foundations}), which constitute the inputs to various tasks. Supported languages can be grouped as \textit{synthetic}, which includes random or established artificial (regular) grammars, Dyck languages or modular arithmetic, to \textit{cognitive}, which includes delayed match-to-sample, conditional one-back 12AX or various non-adjacent dependency (NAD) tasks.

The tool allows for flexible customization of the specific details and parameters of many of these languages. For example, languages with NADs may contain simple long-distance (e.g., in counting tasks),  cross-serial or center embedded dependencies, with many variations regarding the nature and position of the filler/distractor elements (see \prettyref{sec:apps-agl}). In addition, many language generators include functionality for producing sequences with diverse forms of rule violations (deviants).

Regular grammars, a central focus of \verb|SymSeq|, are internally represented using the indexed, Markov-style notation \citep{Warren2015} with symbols in the states and probabilities on the transitions. The index, which is a subscript added to the state label but is not part of the output symbol (e.g., $A_2 \rightarrow A$), is a method to unambiguously distinguish between states with the same symbol occurring in different contexts. This representation also enables tapping into a wide array of Markov analysis tools for studying artificial grammars. Grammar objects can be instantiated in four different ways: inferred from user-provided sequences, loaded from a set of preset grammars, manually specified using a parameter dictionary, and randomly generated based on certain constraints. This latter method is a novel feature of \verb|SymSeq|, enabling users to generate constrained regular grammars with a prescribed level of complexity (via the procedure detailed in the next section), which makes it suitable to systematically evaluate the computational constraints imposed by sequences of varying complexity on biological and artificial models (\prettyref{sec:apps-benchmarks}).

\paragraph[short]{\textbf{Analysis}}
\verb|SymSeq| includes a comprehensive suite of tools for analyzing sequential data at multiple scales of structural organization. Rather than categorizing metrics by methodological approach (distance-based, information-theoretic, statistical), we organize them hierarchically according to the level of linguistic structure they target -- from individual symbols to complete generative grammars. This taxonomy reflects the nested organization inherent to sequential structure and enables systematic investigation of complexity at each level independently. We briefly outline the four hierarchical levels below, with formal definitions, references and computational details provided in \prettyref{sec:metrics}, practical applications illustrated across \prettyref{sec:apps}, and an integrative discussion in \prettyref{sec:discussion}.

\textit{Token-level} metrics focus on individual symbols and their local context, capturing statistical properties such as frequency distributions, n-gram patterns, and temporal persistence. These measures characterize the basic building blocks of sequences but remain agnostic to global structure.

\textit{String-level} metrics evaluate the internal organization, information content, and compressibility of individual sequences. This includes entropy-based measures of unpredictability (Shannon entropy, permutation entropy), algorithmic complexity metrics reflecting compressibility (Lempel-Ziv, effective measure complexity), and linguistic diversity measures. Most operate on isolated strings without requiring corpus-level statistics.

\textit{String-set level} metrics characterize relationships and variability across collections of sequences, enabling corpus-level and comparative analyses. Examples include pairwise distance measures (edit distance, normalized compression distance), information-theoretic quantities (mutual information, string-set entropy), and psycholinguistically motivated metrics such as associative chunk strength that assess sequential predictability based on training exposure.

\textit{Grammar-level} metrics overcome the limitations of finite string samples by characterizing the underlying generative mechanisms themselves. These measures capture the computational capacity and structural principles of the grammar, including topological entropy (quantifying exponential growth in generative capacity), Markov order estimation, hierarchical dependency structure, and production rule complexity. By operating at the grammar level, these metrics provide language-theoretic characterization independent of sampling biases.

This hierarchical organization enables both fine-grained analysis at specific structural scales and systematic comparison across grammars and datasets. The above selection represents a foundational subset drawn from cognitive science, information theory, and formal language theory, designed for extensibility. At the implementation level, analysis functions are modular and can be used independently of the broader toolkit, accepting standard Python data structures (nested lists, NumPy arrays) as input. Where multiple definitions exist in the literature (e.g., variants of associative chunk strength in \citet{Knowlton1996} and \citet{Bailey2008}), \verb|SymSeq| includes all relevant versions with precise documentation of sources and differences.

\paragraph[short]{\textbf{Task targets}}
\verb|SymSeq| distinguishes between two primary types of symbolic sequence processing tasks: \textit{language recognition} and \textit{language transduction}.

Language recognition involves determining whether a given input string conforms to a specific grammar. This is commonly framed as a classification problem -- deciding whether a sequence like ``abba'' belongs to a language defined, for example, by palindromic or nesting rules. However, such tasks can be problematic in practical machine learning contexts because generating negative examples (i.e., strings not in the language) in a principled and exhaustive way is often intractable. As an alternative, recognition is frequently approximated via proxy tasks like next-symbol prediction (e.g., predicting the next character in a well-formed nested sequence like ``a(b(c))'' ) which are easier to implement but introduce issues when sequence length distributions are unknown or variable. Implemented tasks include n-step memorization and prediction at the level of single symbols or longer chunks, along with grammar-specific string legality tests.

Language transduction, on the other hand, shifts focus from classification to functional mapping: given an input sequence from a language $\mathcal{L_\mathrm{inp}}$, produce a corresponding output from a possibly different language $\mathcal{L_\mathrm{out}}$ according to a well-defined transformation. Examples in \verb|SymSeq| include copying a string in reverse (e.g., \ssbquote{abc} $\rightarrow$ \ssbquote{cba}), generating closing brackets for a partial open sequence (e.g., \ssbquote{[((} $\rightarrow$ \ssbquote{))]}) or the odds-first task which requires processing cross-serial dependencies (for a string $s_1s_2s_3s_4s_5s_6$ it should output the symbols with odd indices first $s_1s_3s_5s_2s_4s_6$, e.g., \ssbquote{aababb} $\rightarrow$  \ssbquote{abbaab}). 
These tasks are particularly useful for evaluating a model’s ability to learn deterministic, structure-sensitive mappings between pairs of sequences.

\verb|SymSeq| supports both recognition and transduction tasks, with the former targeting the ability to abstract and generalize grammatical rules, while transduction tasks assess whether a system can implement deterministic, rule-governed transformations. This separation aligns with theoretical distinctions in FLT, while also supporting diverse evaluation strategies across both cognitive modeling and machine learning. Both types include tasks that place varying levels of processing demands on the underlying system.

Despite this conceptual distinction, \verb|SymSeq| follows a ``one-input -- multiple targets'' principle and leverages a modular structure to enable users the specification of multiple tasks and corresponding target outputs for the same language generator. This can be particularly useful in the context of Reservoir Computing \citep{Maass2002}, where it is common to evaluate the same network on many tasks simultaneously. 

\paragraph[short]{\textbf{SeqWrapper}}
Most components of \verb|SymSeq| can be used individually by accessing the relevant functions and classes through standard Python calls. To standardize the entry point and interface to \verb|SeqBench| and other postprocessing pipelines, we implement a lightweight wrapper class \verb|SeqWrapper|, which provides convenient loading routines to instantiate complete datasets and experimental setups based on YAML or Python dictionary parameter configurations. It constructs and exposes language generator objects and creates train/test datasets, which can be pre-generated (offline mode) or created on-the-fly (online mode) by accessing the relevant objects. 

\subsubsection{Complexity-Guided Grammar Synthesis}
\label{sec:grammar-synthesis}
A core feature of \verb|SymSeq| is the generation of structurally parameterized regular grammars with controllable complexity. Users can instantiate fully-specified grammar objects or sample a random grammar based on constraints from a wide range of properties, including but not limited to: alphabet size, level of ambiguity, number of initial and terminal states, transition density. As a primary measure for grammar complexity, we use the information-theoretic Topological Entropy (TE) introduced by \citet{Robinson1998} and applied to AGs by \citet{Bollt2000} (see also \prettyref{sec:metrics}). In these works, an AG is treated as a dynamical system that can generate an infinite number of sequences from a finite set of symbols. Quantifying the intuitive premise that the complexity of a grammar increases with the number of unique strings it can produce, TE is defined as the asymptotic exponential growth rate of the number of grammatical strings as a function of string length. 

\begin{figure*}[h!]
	\centering
    \makebox[\textwidth][c]{
	   \includegraphics[width=180mm]{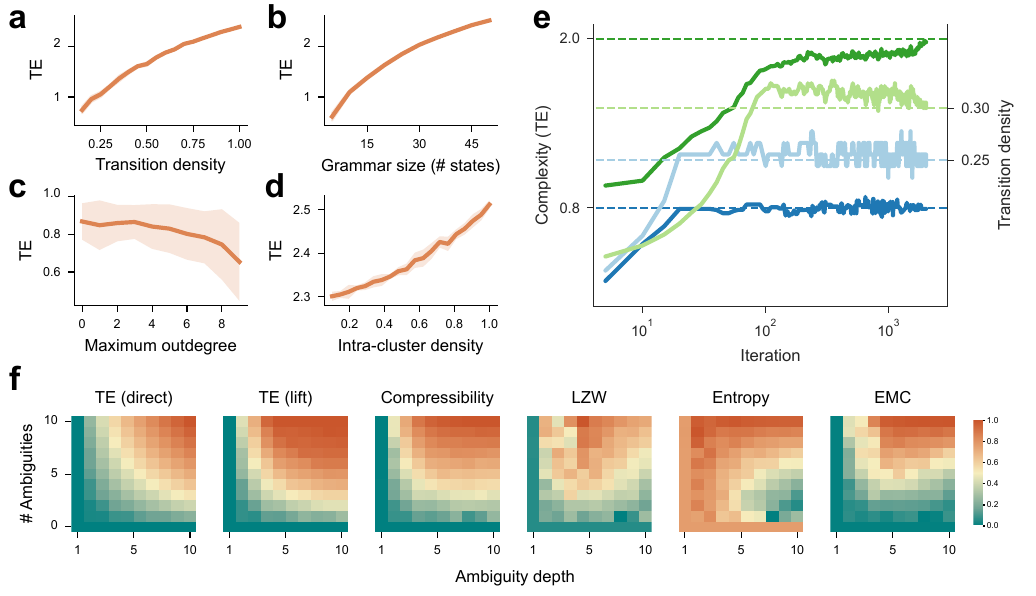}
    }
	\caption[Topological Entropy (TE) as measure of grammar complexity.]{\textbf{Topological entropy as a measure of grammar complexity.} 
	\textbf{a-d)} TE as a function of various grammar properties: \textbf{a)} mean transition density ($|\mathcal{Q}| = 11$); \textbf{b)} grammar size (total number of states in $\mathcal{Q}$, density fixed at $0.25$); \textbf{c)} random grammar with fixed maximum outdegree of one state ($|\mathcal{Q}| = 10$, mean density $0.25$); \textbf{d)} clustering strength ($|\mathcal{Q}| = 100$, $4$ clusters composed of $10$ states, density $0.1$). All results are averaged across 100 trials. 
	\textbf{e)} Convergence of grammar generation with target TE ($0.8$ blue and $2.0$ green, dark shades) and mean transition density ($0.25$ blue and $0.3$ green, light shades).
	\textbf{f)} 
    Comparison of TE and other standard complexity measures, as a function of the number and depth of ambiguities (transition density $0.25$). 
	Non-structural metrics -- Entropy, Compressibility, Lempel-Ziv-Welch (LZW) and Effective Measure Complexity (EMC) -- were computed on 20000 strings drawn randomly from the respective grammars. Results are averaged across 10 different grammars and normalized individually for each metric. Note that the depth parameter is ignored for no ambiguities.
	}
	\label{fig:grammar-synthesis}
\end{figure*}

In practice, the TE of a grammar $\mathcal{G}$ can be obtained by computing the largest real eigenvalue of its topological (boolean) transition matrix (see Methods for details). This \textit{``direct''} approach enables a fast and precise computation of the metric, in contrast to the slower, cumbersome and error-prone \textit{``lifting technique''} proposed initially by \citet{Bollt2000}. For comparison purposes, \verb|SymSeq| provides efficient implementations of both methods.

For grammars satisfying certain mild constraints (see \prettyref{sec:metrics}), understanding how its properties affect the complexity reduces to analyzing how the structure of a binary matrix influences its spectral radius. By the Perron-Frobenius theorem \cite{horn2012matrix}, the spectral radius is bounded from above by the maximum outdegree, although this limit is rarely reached for sparse matrices typically used in AGL studies. In practice, transition density and grammar size are among the most significant determinants of TE (see \prettyref{fig:grammar-synthesis}a,b), consistent with the intuition that a larger set of symbols and more frequent transitions between them yield more complex sequences. To a lesser extent, TE is also impacted by finer structural elements: hub states (here, nodes with high outdegree) can reduce complexity and increase variability across specific instances, particularly for small and low density grammars (see \prettyref{fig:grammar-synthesis}c). This is in contrast with the general expectation that high-degree nodes dominate the principal eigenvector and therefore translate to a larger spectral radius and TE. Other structural features, such as clustering, can also modulate complexity: stronger clustering tends to elevate TE (see \prettyref{fig:grammar-synthesis}d), although the magnitude depends on the number and size of the clusters.

To generate grammars with specific target complexity and priors on various properties (e.g., density, see \prettyref{fig:grammar-synthesis}e), \verb|SymSeq| uses an iterative sampling algorithm that leverages Glauber dynamics on exponential random graph models (ERGMs) to construct a directed graph that satisfies all constraints. 
Starting from an initially random graph of specified size, the algorithm iteratively updates one edge at a time and converges to the stationary distribution encoded by the energy function (Hamiltonian). This scoring function, which incorporates the (weighted) target values of the specified properties, evaluates the fit of the current graph. The approach is efficient for the small and intermediate sized grammars typically used in sequence learning studies, and can be easily extended to account for additional features (e.g., cycles) as long as they can be quantified in a reasonable time.

Compared to other complexity measures, the (direct) TE method with subscripts used here is the most sensitive and consistent with the grammar complexity as a function of ambiguity (\prettyref{fig:grammar-synthesis}f). Importantly, TE is \textit{structural but blind to bias}: although it does capture combinatorial growth of valid strings, it is insensitive to the actual symbol identities (reflected by the symmetry in \prettyref{fig:grammar-synthesis}f) as well as any structure engraved in transition probabilities. Compression-based metrics exhibit a similar overall pattern, but compressibility (compression ratio) emphasizes the number of ambiguous states more than their repetitions, whereas LZW \citep{Welch1984} does not accurately reflect the increase in complexity with ambiguity depth. In contrast, information-theoretic measures of unpredictability, such as Shannon entropy, display an opposite and more nonlinear behavior: for a small number of ambiguities entropy actually decreases with repeated occurrences, but otherwise it shows minimal variation with the number of ambiguities.
Effective Measure Complexity \citep[EMC,][]{Grassberger1986}, which quantifies the balance between predictability and surprise in a sequence, exhibits a pattern similar to LZW but with a more graded response. For a given number of ambiguities, EMC shows a bell-shaped dependence on repetition depth -- low for small depth, peaking at intermediate levels, and declining again for many repetitions. Thus, such measures have difficulty capturing subtle contextual variations in sequences and are not well-suited for systematic control. 

Using the direct TE method and iteratively sampling the grammar space allows \verb|SymSeq| to generate regular grammars of prescribed complexity, thus enabling systematic navigation of the complexity landscape to design controlled experimental paradigms, construct graded benchmark suites, and investigate structure-learning relationships across architectures.

\subsection{\texttt{SeqBench}: Token embeddings and datasets}

\begin{figure}[!ht]
    \centering
    \includegraphics[width=0.9\linewidth]{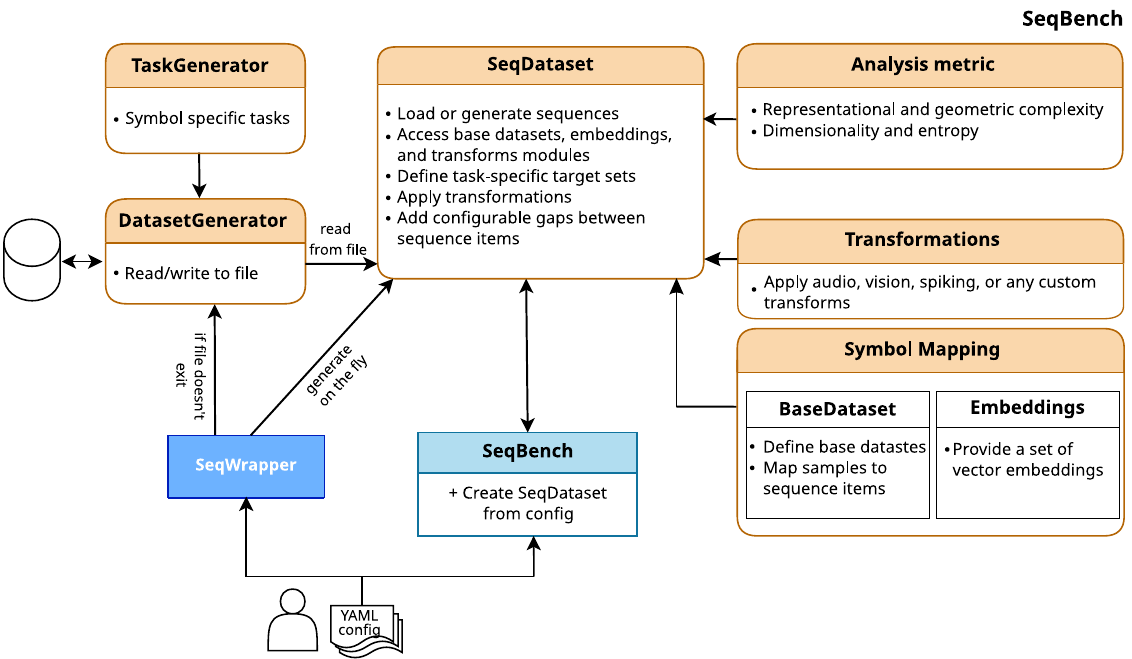}
    \cprotect\caption{\textbf{Schematic overview of \texttt{SeqBench}.} \verb|SeqBench| provides a modular pipeline for generating, transforming, and analyzing symbolic sequence datasets. A user-defined configuration is passed to the SeqWrapper, which serves as the interface to the \verb|SymSeq| sequence generator. The DatasetGenerator either loads existing symbolic data from disk or invokes the SeqWrapper to create new sequences. The SeqDataset forms the core of the system, combining symbolic sequences with base datasets or embeddings, applying perturbations such as gaps and noise, and supporting optional transformation modules for audio, vision, generic tensor operations, or spiking representations. Symbol Mapping defines how tokens correspond to dataset samples or vector embeddings, while the Analysis module computes complexity metrics, including dimensionality and entropy.
    }
    \label{fig:seq_bench_overview}
\end{figure}

The \verb|SeqBench| package offers a versatile pipeline for transforming abstract symbolic sequences into concrete, task-ready datasets, with fine-grained control over both the structural complexity of sequences (inherited from \verb|SymSeq|), input representations, and symbol semantics (grounding) (\prettyref{fig:seq_bench_overview}).
Its core functionality lies in the flexible mapping between symbolic sequences and embedded representations. Users can define custom symbol-to-stimulus mappings via base datasets (e.g., mapping symbols to images, audio samples, or arbitrary objects), specify vector embeddings of controlled dimensionality and geometric structure, or combine both approaches. This enables seamless adaptation from simple one-hot encodings for theoretical analyses to naturalistic, multimodal stimuli for cognitive experiments or neural network training.

Beyond basic encoding, \verb|SeqBench| supports a wide range of custom transformations of the embeddings, including domain-specific operations for audio (spectral processing, noise injection) and vision (geometric distortions, color manipulations), and allows the controlled induction of temporal perturbations, including configurable gaps between sequence items, temporal jitter, and presentation timing manipulations. The framework also includes tools to quantify embedding complexity through representational metrics (dimensionality, effective rank) and geometric measures (pairwise distances, manifold structure).

This modular design enables straightforward data augmentation, ablation studies across representational modalities, and task-specific adaptations while maintaining independent control over rule-based symbolic structure and input characteristics. To streamline usage, \verb|SeqBench| can generate raw sequences on-the-fly via callbacks to \verb|SymSeq|, load stored precomputed sequences, or import external datasets, with complete experimental pipelines specified through human-readable YAML configuration files accessible via the \verb|SeqWrapper| interface.

\subsubsection{Main components}

\paragraph[short]{SeqDataset} inherits from \verb|torch.utils.data.Dataset| and is the central dataset interface of \verb|SeqBench|. Depending on the configuration, sequences are either read from pre-generated files or produced on-the-fly via the \verb|SymSeq| sequence generator (see DatasetGenerator).
Raw sequences are buffered in CPU memory. Sequence items are then mapped to dataset samples, or the specified embedding (see Symbol mapping). After these mappings, they are assembled into complete training examples that contain both input and target sets.
We provide dedicated modules to define target sets, for example, predicting the set of possible next items in the sequence.
The SeqDataset also supports configurable gaps between sequence items and other perturbations that are controlled by the configuration dictionary, which specifies probabilities, durations, gap lengths, and noise magnitudes. These operations are applied within the \texttt{\_\_getitem\_\_} method, so the same base data can be reused across experiments.

\paragraph[short]{DatasetGenerator}
manages how symbolic sequences are created, serialized, and restored. It serves as the interface between the \verb|SymSeq|-based sequence generator and the storage format used in \verb|SeqBench|.
The generation can be performed either in a single process or in parallel across multiple processes to accelerate dataset creation, with each sample written directly to file in a consistent textual format. Each sample contains the class sequence, the corresponding state sequence, and its length.
The generator also records the transition matrix of the underlying Markov chain and writes it to a dedicated file, ensuring that the probabilistic dynamics can be replayed or inspected after generation. A YAML configuration file is written alongside the data, capturing the parameters used in the run and allowing reproducibility.

\begin{figure}[!ht]
    \centering
    \makebox[\textwidth][c]{
        \includegraphics[width=180mm]{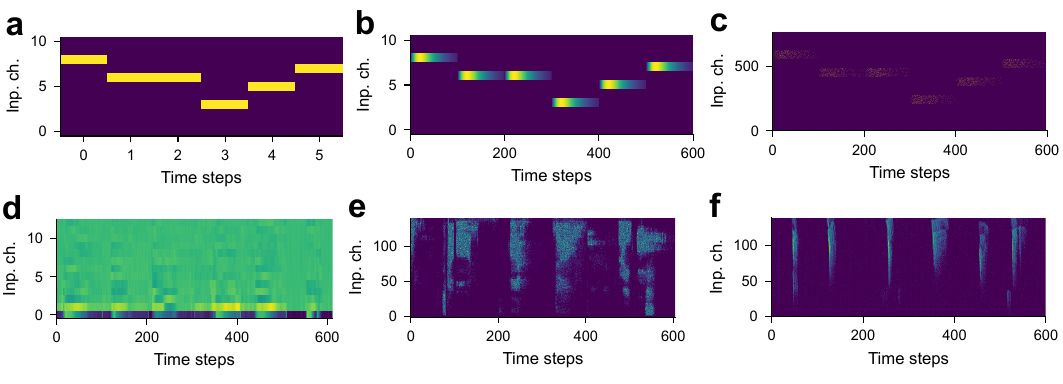}
    }
    \caption{\textbf{Example sequence encodings produced within \texttt{SeqBench}.} (a) Discrete token embedding using one-hot vectors. (b) Embedded sequence filtered with an exponential kernel and extended across input channels. (c) Rate-based Poisson spike trains derived from the embedded representation. Encoded continuous signals \textbf{(b)} are considered the mean density (instantaneous firing rate) of a series of independent Poisson processes to translate the signals into spike trains. (d) Google Speech Commands (GSC) representation, where each token is mapped onto one individual spoken command instance. (e) Rate-based spike encoding of GSC. (f) Spiking Speech Commands dataset representation, where each token is mapped to an instance of a spike-coded speech command.
    }
    \label{fig:symbol_mapping}
\end{figure}

\paragraph[short]{Symbol mappings} specify how each sequence token is embedded and how abstract symbolic sequences are mapped onto dynamic, numerical data structures that have a defined meaning.  

The \textbf{BaseDataset} class defines the core interface for mapping abstract symbolic tokens to concrete samples. Any classification dataset -- whether natively supported by SeqBench, loaded through the Tonic backend \cite{Lenz21_5079802}, or provided by user-integrated backends -- can serve as a source for embedding symbolic sequences. To enable this mapping, BaseDataset maintains an internal index structure that links each class label to all samples belonging to that class. When \verb|SeqBench| encounters a symbolic token, it resolves the token to its corresponding class and draws a specific sample from the appropriate index set. This mapping is constructed once during SeqDataset initialization by scanning the dataset and is subsequently stored for efficient reuse across sequence generation.

Beyond dataset-based mappings, \verb|SeqBench| supports \textbf{vector embeddings} that provide static representations of symbolic tokens through methods such as one-hot encodings, binary codewords, random vectors, or custom embedding functions. These lightweight alternatives enable rapid prototyping and theoretical investigations without requiring external datasets.

\textbf{Transformations} provide a flexible interface to modify or extend embeddings produced by either datasets or embedding modules. Users can compose arbitrary transformation functions to adapt data representations to specific tasks and modalities without altering the core dataset logic. This interface supports operations at multiple stages, including raw samples, embedded representations, or temporally structured data, making it straightforward to implement custom preprocessing pipelines or alternative coding schemes for downstream experiments. Available transformations span multiple modalities: audio operations (mel features, MFCCs, power law compression, adaptive normalization), vision preprocessing (standard image transforms), generic tensor manipulations (convolutions, dimension operations), and spiking neural network encodings (temporal binning, Poisson spike generation, rate coding).

Together, these components provide a modular framework that decouples symbolic structure from concrete instantiation, enabling systematic control over sequence complexity, input representation, and stimulus grounding as independent experimental dimensions. \verb|SeqBench| thus bridges formal sequence generation via \verb|SymSeq| with flexible dataset construction, supporting applications ranging from controlled cognitive experiments with naturalistic stimuli to systematic benchmark suites for evaluating sequential processing in neural networks.

\newpage
\section{Applications and use-cases}
\label{sec:apps}

\subsection{Psycholinguistics and cognitive psychology: designing experimental paradigms and datasets}
\label{sec:apps-agl}
Experimental paradigms in psychology and cognitive neuroscience often appear distinct, designed to probe particular aspects of cognition such as memory, attention or language. Yet beneath this apparent diversity lies a common structure that can be described using the same symbolic framework. Tasks like delayed match-to-sample, 12AX, odd-ball paradigms, serial reaction times or sequence recall all define structured relations between temporally or contextually separated symbols, often captured by simple production rules. Despite targeting different cognitive domains, they depend on similar processes of maintaining, updating, and binding representations across intervening material. However, experimental protocols and benchmark datasets in behavioral research and computational modeling are often tailored to a single task or domain, making it difficult to compare models or results across paradigms. \verb|SymSeqBench| leverages the shared symbolic structure to address this gap, by using a declarative, grammar-based format that lets researchers generate, analyze and compare datasets across different domains using the same representational framework. This unified approach makes it easier to systematically explore how task complexity affects performance, how learning transfers between tasks, and how models generalize across paradigms, offering a principled way to identify computational mechanisms that connect cognitive phenomena that might otherwise seem unrelated. Whether the objective is to exploit these commonalities or to generate constrained experimental protocols and datasets, \verb|SymSeqBench| provides the necessary functionality to cover a wide range of cognitive tasks.

\subsubsection{Non-adjacent dependencies and temporal binding}
\label{sec:apps-nad}
One powerful conceptual bridge across perceptual learning, psycholinguistics, working memory, and executive control are non-adjacent dependencies (NADs), which involve maintaining and integrating information that is separated in time, space, or through noise. As illustrated by the stereotypical A-X-B structure, a NAD occurs when two (or more, dependent) elements (A and B) must be related or integrated while divided by intervening, potentially distracting elements (X, fillers).  

\begin{figure}[t]
    \centering
    \makebox[\textwidth][c]{
        \includegraphics[width=180mm]{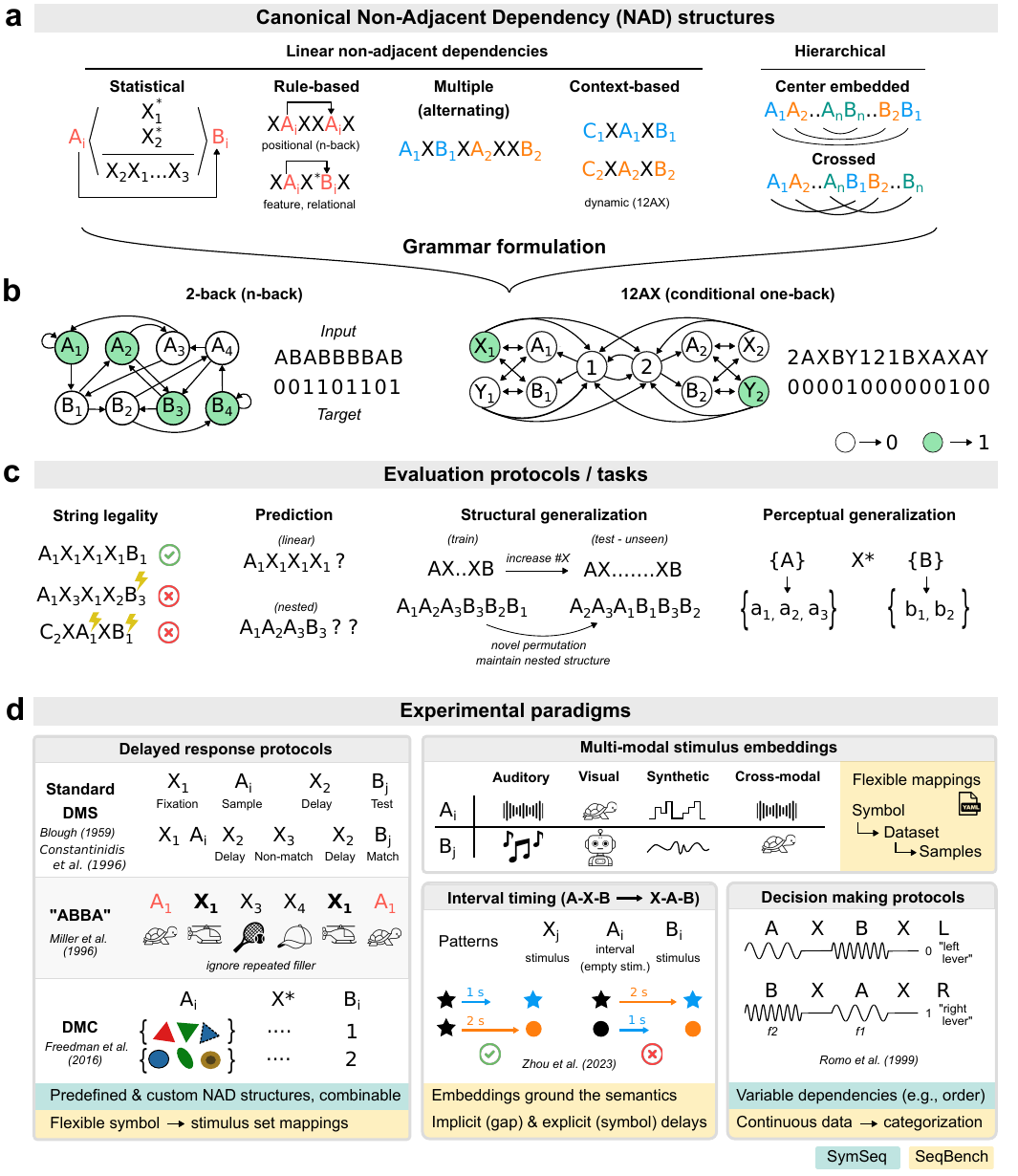}
    }
    \cprotect\caption{\textbf{Non-Adjacent Dependencies (NADs) as a unifying template across cognitive domains.}
    Many seemingly distinct cognitive tasks can be reduced to a single underlying principle: maintaining and binding non-adjacent elements in a sequence. \verb|SymSeqBench| enables seamless implementation of these paradigms by parametrizing the sequence, dependency rule, and context, as well as the symbol mappings and embeddings.
    \textbf{(a)} Canonical NAD structures may include linear (e.g., statistical) and hierarchical dependencies (e.g., crossed).
    \textbf{(b)} Although these typically require careful experimental design for better control, many of them can be formulated and represented as regular grammars in conjunction with appropriate target labels. For simplicity, some formal notations (e.g., start and end states) are not depicted.
    \textbf{(c)} Computationally, NAD structures can be evaluated in a variety of ways, including string legality, prediction, or structural and perceptual generalization.
    \textbf{(d)} Examples of psychological paradigms expressible as NAD structures, which can be explored using our framework through flexible configurations of \verb|SymSeq| (teal boxes), \verb|SeqBench| (yellow boxes), or a combination of both. These include but are not limited to Delayed Match-to-Sample (DMS) and Delayed Match-to-Category (DMC) protocols, multi- and cross-modal experiments with different symbol - dataset - sample mappings, learning of interval timings through symbol - silent stimulus mappings, and (perceptual) decision making tasks requiring a rule-based stimulus comparison and action selection.
}
    \label{fig:nad}
\end{figure}

The specific NAD structure form can range from simple, linear dependencies to more complex, hierarchical ones (\prettyref{fig:nad}a). Linear statistical NADs involve surface co-occurrence learning based on probabilistic associations between distant elements \citep{gomez_variability_2002}, while rule-based NADs \citep{wallis_single_2001, kane_working_2007} go beyond this by encoding abstract category-to-category mappings that generalize to novel items, requiring symbolic binding and working-memory maintenance. Multiple or alternating NADs \citep{wilson_non-adjacent_2020,deocampo_concurrent_2019,wang_learning_2018} interleave several dependency types within a sequence, increasing interference and taxing parallel rule tracking. Contextual NADs \citep{wallis_single_2001,Mante2013} (e.g., 12AX), add another layer of complexity by making the relevant dependency conditional on a preceding cue, thus demanding dynamic context updating and executive control. Finally, hierarchical or center-embedded NADs \citep{Vries2012} introduce genuine structural recursion, where dependencies are nested within one another, imposing heavy syntactic and memory loads. Across this continuum, complexity grows from purely statistical tracking to context-dependent rule selection and ultimately to hierarchically embedded representations that require compositional structure and recursive processing.

While these structures can, in principle, be expressed through formal grammars (\prettyref{fig:nad}b), \verb|SymSeq| implements most of the NAD generators and related tasks as customizable modules (instead of explicit grammars) for several reasons: dedicated modules allow more intuitive control of desired statistics and experimental design; specifying complex grammars can be impractical; and \verb|SymSeq| currently provides explicit support only for regular grammars. The \emph{n-back} task, for instance, can benefit from a low match probability \citep[e.g.,][]{scharinger_comparison_2017} to avoid developing a biased "match" response, the introduction of lures (near misses) and repetitions for fine-grained complexity control \citep{kane_working_2007}, and a homogeneous distribution of matches to counter primacy/recency and serial position effects \citep{juvina2007modeling}.

An agent’s ability to handle such structures can be tested in different ways (\prettyref{fig:nad}c), most directly by assessing whether it can recognize string legality from statistical, rule-based or contextual dependencies. Other approaches include evaluating next-item(s) prediction, syntactic generalization to novel sequence lengths or structure-preserving compositions, and perceptual generalization across variations that preserve the underlying dependencies. 
 
Crucially, many well-known experimental paradigms can be reinterpreted or reformulated as variants of NAD tasks and are therefore easily implementable in \verb|SymSeqBench|, as illustrated schematically in \prettyref{fig:nad}d. In delayed response protocols \citep{blough_delayed_1959,constantinidis_neuronal_1996, freedman_neuronal_2016}, the sample ($A_i$) and match ($B_i$) items serve as dependent elements, while other stimuli possibly presented during fixation or delay periods act as fillers (X) \citep{miller_neural_1996}.  Through flexible symbol mappings in \verb|SeqBench| -- each symbol can be independently mapped to multiple samples from distinct datasets --, users can generate both arbitrary experimental protocols (e.g., delayed match-to-category) and scenarios involving multimodal datasets.

\FloatBarrier

For example, interval timing experiments \citep{zhou_multiplexing_2023} can be modeled as an X-A-B structure, where the cue $A_i$ is the delay/interval itself and determines the associated item $B_i$. In \verb|SeqBench|, delays can be expressed implicitly (dedicated inter-stimulus-gap parameter) or explicitly (mapping symbols to empty/silent stimuli). This flexibility in the NAD structure also supports decision-making protocols \citep{romo_neuronal_1999}, for instance, using synthetic datasets that combine continuous-time and discrete inputs at the embedding level. 

This section presented only a high-level overview NADs, intended to illustrate the links between computational and experimental paradigms. While all of these approaches are supported in \verb|SymSeqBench|, for clarity, we omitted specific implementation details here and refer the user to the extensive documentation accompanying the tool.

\subsubsection{AGL dataset generation}

NADs have classically been studied in the context of working memory and attention experiments, because detecting and maintaining non-local relations in a sequence places demands on these cognitive systems. By contrast, artificial grammar learning (AGL) represents a broader experimental framework for investigating how humans (as well as computational models and other mammals) learn sequential dependencies of various kinds -- adjacent or non-adjacent, linear or hierarchical -- through mere exposure to structured inputs. A typical AGL study involves exposing subjects to \emph{grammatical} (G) strings obeying certain rules that are to be acquired (often) implicitly, and later require them to distinguish novel G strings from non-grammatical (NG) ones. Beyond the complexity of the rule-encoding grammar itself \citep{schiff_does_2014}, the structural properties of the stimulus (string) sets can be tailored to control a range of psychologically relevant factors in order to manipulate task difficulty and to probe specific learning theories \citep{vokey_salience_1992,Knowlton1996,meulemans_implicit_2003}.

Constructing balanced datasets that account for many factors simultaneously is a delicate business and has traditionally required hand-picking stimulus sets refined through iterative trial-and-error. However, such approaches do not scale well for complex grammars, larger datasets, or more than a handful controlled properties. Existing software typically focuses on the a posteriori analysis of such datasets \citep{cook_agsuite_2017, pyagl}, and only a few tools -- such as \verb|RNNExploration4SymbolicTS| \citep{cahuantzi_comparison_2021} and \verb|AGL StimSelect| \citep{Bailey2008} -- were released as packages to automate dataset generation. Of these, only \verb|AGL StimSelect| incorporates psychologically relevant factors and yields experiment-ready datasets, but it was developed in MATLAB (a proprietary environment) and is no longer maintained. Building on its core concepts while greatly extending functionality, \verb|SymSeq| fills a key gap by providing an actively maintained, open-source framework for generating flexible and well-controlled AGL datasets.

\begin{figure}[!ht]
    \centering
    \includegraphics[width=1.0\linewidth]{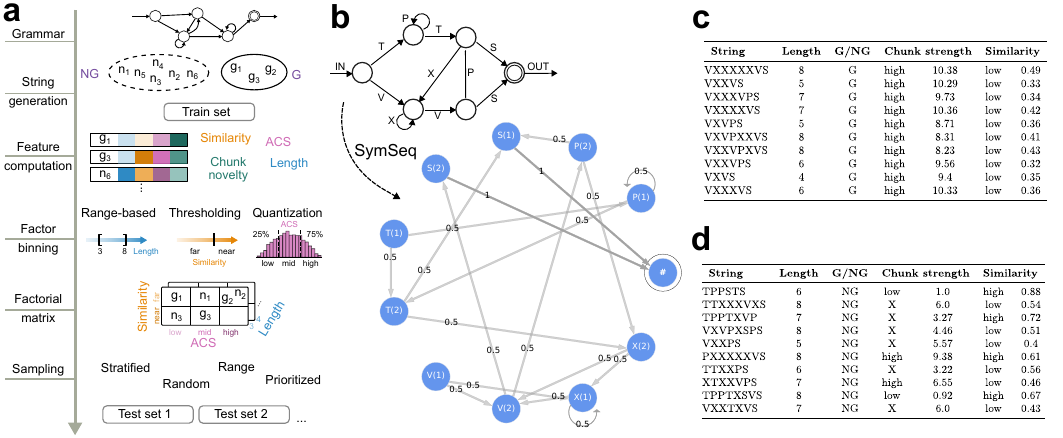}
    \cprotect\caption{\textbf{Generation of AGL datasets using \texttt{SymSeq}.} 
    With a few lines of code, users can generate training and test strings based on a range of constraints and complexity measures from preset or custom grammars. 
    \textbf{a)} Schematic of the factorial sampling procedure, which begins with the generation of set of G and NG candidate strings from which a training set is selected. User-specified features are then computed for all remaining candidates, and continuous features are binned into discrete factor levels (e.g., via quantiles or thresholds). Candidate strings are subsequently filtered according to factor-level specifications, and final test items are drawn using one of several sampling procedures implemented (e.g., stratified or range-restricted).
    \textbf{b)} Graphical representations of the Reber grammar in AGL style (upper left) and Markov-style used in \verb|SymSeq| (lower right).
    \textbf{(c-d)} Example datasets using 16 training strings, with factors including grammaticality, global ACS \citep{Knowlton1996} and similarity (to training strings), and string length restricted between 3 and 8.
    \textbf{c)} Set of grammatical test strings with high chunk strength ($\leq 25$th or  $\geq 75$th quantile) and low train-set similarity ($< 0.5$). 
    \textbf{d)} Set of non-grammatical test strings with no chunk-strength restrictions and including both low and high similarity.
    }
    \label{fig:agl-datasets}
\end{figure}

Users can specify a wide range of constraints and factors, from the number of grammatical and ungrammatical strings to the inclusion of specific deviations or positional exclusions (\prettyref{fig:agl-datasets}a). Supported \emph{factors} include all quantitative metrics available within \verb|SymSeq|'s analysis module, with \emph{levels} individually definable for each factor or derived automatically using multi-bin quantile partitioning. Initially, a variable number of G and a larger pool of NG candidate strings is produced, after which all relevant factors are computed in a parallelized manner to ensure efficiency even for large-scale datasets. A heuristic selection stage then samples specific \emph{factor cells} according to the user’s experimental design and stratification requirements. While this step allows for fine-grained control over the factor structure of the resulting dataset, it can face limitations when strict constraints prevent sufficient sample generation.

The resulting datasets (see, for example, \prettyref{fig:agl-datasets}b-d), including the computed factor structures, are returned as pandas dataframes for seamless downstream integration. Compared to \verb|AGL StimSelect|, \verb|SymSeq| adopts a more flexible and factor-driven approach. Whereas the former iteratively constructs training sets in a controlled and incremental manner, \verb|SymSeq| emphasizes exploratory and large-scale generation, supporting a broader range of metrics and stratified sampling schemes. This design makes it particularly suited for high-throughput experimentation, model benchmarking, and systematic comparisons across AGL paradigms.

\subsection{Cognitive computing benchmarks: evaluating biological, neuromorphic and machine learning architectures on psychologically meaningful tasks}\label{sec:apps-benchmarks}

Understanding how different neural systems process and learn sequences is central to both neuroscience and artificial intelligence. Biologically detailed models aim to uncover the circuit and cellular-level mechanisms that give rise to temporal and contextual processing in the brain. Neuromorphic implementations, in turn, translate these principles into efficient, real-time computation under physical and energy constraints. The following sections demonstrate how \verb|SymSeqBench| provides a systematic benchmarking framework for both domains: first by evaluating biologically inspired models trained with local learning rules, and then by assessing neuromorphic architectures designed for low-power temporal processing.

\subsubsection{Biological neural networks}
Biologically-detailed models of sequence learning, such as recurrent networks of spiking neurons equipped with local or unsupervised learning rules, are typically developed to capture 
the transition and timing aspects \citep{murray_learning_2017, cone_learning_2021,  bouhadjar_sequence_2022,Bouhadjar2023,leugering_dendritic_2023,duarteDynamicStabilitySequential2014}, with fewer propositions addressing chunking \citep{fonollosa_learning_2015, asabuki_neural_2022}, artificial grammar learning \citep{Duarte2014a}, symbolic sequence encoding \citep{Duarte_Uhlmann_Broek_Fitz_Petersson_Hagoort_Morrison_2017}, or hierarchical structure transfer \citep{Zajzon2018}. Recent work has also shown that spike-rate adaptation can support working memory demands in language processing \citep{Fitz2020} and that simplified dendritic morphologies preserve essential memory-expanding transformations \citep{quaresimaTripodNeuronMinimal2023} suitable for sequential learning. While these models provide insights into the biophysical and circuit-level mechanisms supporting cognitive computations, they are often evaluated on a narrow range of handcrafted tasks with limited input variation, intended to highlight a single functionality. As a result, their computational scope and limitations remain difficult to assess methodically, and direct comparisons across models are rare. A case in point is the ability to process higher-order (non-)Markovian sequences \citep{maes_learning_2021,cone_learning_2021,asabuki_neural_2022, bouhadjar_sequence_2022}, which is typically evaluated on a few randomly and manually selected sequences with fairly static, synthetic stimulus encodings (e.g., Poisson-rate spikes), seldom considering essential factors such as sequence complexity, sensitivity to deviants or symbol repetitions, or broader variability in stimulus statistics.

\verb|SymSeqBench| facilitates experiments beyond prototypical demonstrations by offering a unified, systematically parameterized suite of sequence-processing tasks that vary structural dependencies, timing, and stimulus complexity, enabling rigorous model-to-model and model-to-data comparisons. The tool enables automatic string generation of tunable complexity that can be further constrained by experimentally grounded metrics, allows flexible switching between synthetic and real input encodings as well as their easy manipulation, and provides fine-grained control over temporal stimulus properties such as duration, inter-stimulus intervals, and noise. 

\begin{figure}[!ht]
    \centering
    \includegraphics[width=1.0\linewidth]{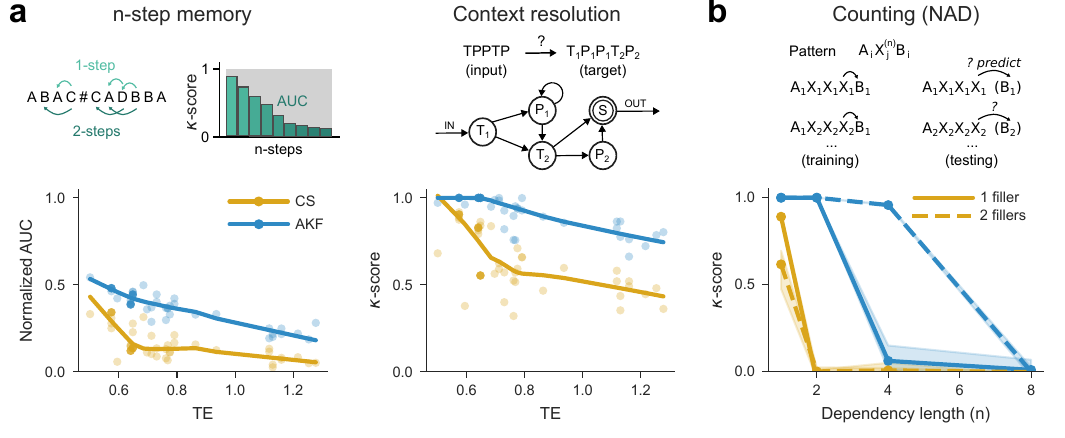}
    \cprotect\caption{\textbf{Performance of biologically-plausible models on cognitive sequence processing tasks.} Two models featuring non-supervised learning -- \citet[][red]{cone_learning_2021} and \citet[][purple]{asabuki_neural_2022} -- were evaluated on memory-related tasks using reservoir computing (see Methods for details).
    \textbf{a)} Learning ambiguous adjacencies in regular grammars created using \verb|SymSeq|, shown as a function of complexity: (left) aggregated n-step symbol (label) memorization (normalized area under the curve (AUC) of Cohen's $\kappa$ across valid n-steps, here 9); (right) context resolution (classification of the indexed state, e.g., $A_1$ or $C_3$). For each property combination ($|\Sigma| = 5$, $|q_0| = 2$, $\mathrm{len}(S_i) \in [2, 10]$,  $\leq 3$ ambiguities with depth $\leq 4$), three random grammars were generated. Each data point represents such a grammar, with curves showing nonparametric LOWESS reggression fits.
    \textbf{b)} Non-adjacent dependency (counting) task with two dependent elements ($A_1B_1$ and $A_2B_2$) and one (solid curves) or two (dashed curves) distinct filler items. Performance is plotted against filler repetitions, based on the prediction accuracy of the final (dependent) element.
    }
    \label{fig:app-bio-nn}
\end{figure}

To illustrate some of these capabilities, we compared two models by \citet[][CS]{cone_learning_2021} and \citet[][AKF]{asabuki_neural_2022} on three example tasks assessing memory capabilities. 
First, \verb|SymSeq|'s AG generator constructs grammars of varying complexity by manipulating the number and repetitions of ambiguous states. 
For each grammar, we generate a set of input strings using symbols from the alphabet, and define two tasks and corresponding sets of target labels: (i) n-step memory ($n \leq 10$), in which the n-th previous symbol must be recalled, and (ii) context resolution, in which not just the current symbol but rather the underlying indexed state must be identified. Identifying the exact state in the grammar is equivalent to disambiguating the relevant history and, in such regular grammars, it can be considered as a proxy for prediction (the indexed representation is Markovian, i.e., memoryless). Performance on both tasks generally declines with increasing TE complexity (see \prettyref{fig:app-bio-nn}a), but the AKF model consistently outperforms CS, possibly reflecting crucial computational benefits of its dendritic processing and nonlinear gating mechanisms.

The third task probes the system's working memory using sequences of the form $A_iX^*B_i$,  a classical formulation of non-adjacent dependencies. Also known as the \textit{counting} task \citep{Lazar2009, Duarte2014a}, this involves correctly predicting the last dependent element $B_i$ after seeing the final filler item $X_j$. While the CS model is unable to handle more than one filler occurrences, AKF can solve the task up to a dependency length of $8$. More interestingly, having two distinct filler items ($X_1$ and $X_2$) significantly improves the model's performance. This is in line with experimental findings in Humans, which demonstrate that larger filler variability can be computationally beneficial \citep{Onnis2003}.

\subsubsection{Artificial and neuromorphic neural networks}

Neuromorphic systems are designed to learn and process data in real time while operating under strict energy constraints.
The input they receive is often sequential and heterogeneous, combining spatial and temporal patterns of varying complexities and scales. 
Many of these systems are still under active development, exploring novel mechanisms that demand rapid iteration to uncover their strengths and limitations.
Designing dedicated benchmarks is therefore becoming increasingly critical \cite{Yik25_neurobench}. Current datasets are either too simple to capture long-term temporal dependencies, such as the Spiking Heidelberg Digits (SHD), Spiking Speech Commands (SSC) \citep{SHD}, or the NeuroMorse \citep{Walters25_NeuroMorse} datasets;
others are narrowly specialized to evaluate particular capacities, such as memorization \citep{Tay21_LongRangeArena};
and datasets such as \citep{Kamradt23_nihs, Hsieh24_ruler} are designed for large-scale networks that are unsuitable for energy-constrained settings.
Recent efforts, such as the Neuromorphic Sequential Arena (NSA), address this gap by providing a suite of real-world temporal processing tasks that evaluate spiking models across performance, efficiency, and scalability dimensions \citep{Chen25_NSA}.
However, NSA lacks explicit control over the complexity of temporal dependencies, limiting its utility for systematically probing models.
The DVS-Gesture-Chain (DVS-GC) introduced in \cite{Vicente25_SeqDVSGesture} constructs a temporal recognition task by forming sequences of gestures, where each sample is built by concatenating multiple individual gestures from the DVS Gesture dataset \cite{Amir17_DVSGesture_CVPR}.
Given that the samples are concatenated randomly, it is difficult to assess or tune the complexity of the gesture chain.
\par
\verb|SymSeqBench| generates sequences with controllable spatial and temporal complexity. Spatial complexity is modulated by selecting a base dataset and concatenating samples into sequences through grammar-based compositional rules (see \prettyref{fig:symbol_mapping}). These rules govern how elements are combined, thereby introducing temporal dependencies at both short and long timescales. Although synthetically generated, these tasks capture essential characteristics of real-world problems, making them both interpretable and relevant for practical applications. This controllability directly addresses the lack of adjustable temporal scales in benchmarks such as NSA, enabling systematic evaluation of how models cope with varying degrees of temporal dependencies.
\par
To demonstrate these properties in practice, we consider representative \verb|SymSeqBench| configurations with the AG grammar settings specified in \prettyref{tab:snn_performance} and \prettyref{tab:ann_performance} using SHD, SSC, and GSC as base datasets.
SHD and SSC samples are processed using a $10\ms$ temporal binning resolution, with their spiking inputs spatially reduced from 700 to 140 channels.
GSC inputs are binned at $10\ms$ and transformed using a 40-channel Mel filterbank.
The neuromorphic baselines are implemented as spiking neural networks composed of 5 fully connected layers of 256 leaky integrate-and-fire (LIF) or adaptive LIF (adLIF) neurons (for more details, see methods and the study by \citet{Bittar22}).
We first discuss these SNN baselines before turning to the ANN baselines (GRU \citep{Cho14_GRU}, Mamba \citep{Gu24_mamba}, Transformer \citep{Vaswani17_attention}), which are provided as reference points to contextualize the performance of SNN models against widely used sequence learning architectures.
We benchmark the models on classifying the indexed states underlying the input sequences (see \prettyref{sec:symseq-main}).
Across the sequences composed of SSC and SHD as base datasets, adLIF networks outperform their LIF counterparts (see \prettyref{tab:snn_performance}), indicating that adaptive neuronal dynamics can be beneficial for learning complex spatio-temporal dependencies \citep{Baronig25_EFadLIF}.
For sequences with the GSC base dataset, the adLIF exhibits reduced performance compared to LIF. Despite being more computationally expressive, the vanilla adLIF used in this study is known to suffer from training instabilities, as outlined in \cite{Baronig25_EFadLIF}.
We leave for future work the improvement of the adLIF model used in the current study by incorporating features discussed in \cite{Baronig25_EFadLIF} and \cite{Fabre25_SiLIF}.
Overall, the performance gap to the theoretical upper bound indicates substantial room for further improvement in model architectures and training strategies.
In contrast, the ANN baselines obtain higher accuracies overall (see \prettyref{tab:ann_performance}). For sequences with lower complexity (TE = 1.71), we use 4-layer architectures with 256 units per layer. In this regime, all three architectures perform well, with the GRU achieving the highest accuracy, suggesting that moderate temporal dependencies can already be captured effectively by classical recurrent models. For sequences with higher complexity (TE = 2.61), we use larger networks with 8 layers of 1024 units. In this setting, Mamba attains the best performance, indicating that its state space formulation scales favorably as sequences become longer and more complex.
Nevertheless, even the strongest ANN baselines exhibit a noticeable performance gap at higher sequence complexity, indicating the need for further development of architectures capable of handling such sequences.
\par
In summary, \verb|SymSeqBench| enables systematic evaluation of sequential models by providing explicit control over spatial and temporal complexity. It reveals clear benefits of adaptive spiking dynamics while exposing persistent performance gaps as sequence complexity increases, even for strong ANN baselines. This makes SymSeqBench a useful diagnostic tool for identifying limitations in current architectures and guiding the development of more effective temporal learning mechanisms.

\begin{table}[h]
\small
\centering
\begin{tabular}{
l |
cc | cc | cc
}
\hline
 & \multicolumn{2}{c|}{Seq SHD}
 & \multicolumn{2}{c|}{Seq SSC}
 & \multicolumn{2}{c}{Seq GSC} \\
 
 & \multicolumn{2}{c|}{TE = 1.35}
 & \multicolumn{2}{c|}{TE = 1.35}
 & \multicolumn{2}{c}{TE = 1.35} \\
 
 & LIF & adLIF
 & LIF & adLIF
 & LIF & adLIF \\
\hline

Nb. param.
 & 327k & 332k
 & 327k & 332k
 & 295k & 299k \\

Accuracy
 & 56.80\% $\pm$ 1.39 & \textbf{65.79\% $\pm$ 1.44}         
 & 49.92\% $\pm$ 0.48 & \textbf{57.44\% $\pm$ 0.6}          
 & \textbf{61.46\% $\pm$ 0.28} & 59.70\% $\pm$ 4.84 \\      

\hline
\end{tabular}
\caption{
Neuromorphic network performance on the context resolution task using temporal SHD, SSC, and GSC as base datasets (abbreviated as Seq SHD, Seq SSC, and Seq GSC). A single gap element is inserted between sequence items, and accuracy is measured during this gap interval. Experimental configuration: alphabet size $|\Sigma| = 11$, $10$ ambiguities with depth $10$, transition density $= 0.04$, $|q_0| = 4$ wit a resulting sequence complexity of $\mathrm{TE} = 1.35$, minimum and maximum sequence length: $\mathrm{len}(S_i) \in [1, 30]$. Results are reported as mean $\pm$ standard deviation over 4 runs.
}
\label{tab:snn_performance}
\end{table}

\begin{table}[h]
\small
\centering
\begin{tabular}{
l |
ccc | ccc
}
\hline
 & \multicolumn{3}{c|}{Seq GSC}
 & \multicolumn{3}{c}{Seq GSC} \\

 & \multicolumn{3}{c|}{TE = 1.71}
 & \multicolumn{3}{c}{TE = 2.61} \\
 
 & Attention & GRU & Mamba
 & Attention & GRU & Mamba \\
\hline

Nb. param.
 & 4.276M & 1.646M & 2.113M
 & 60.041M & 61.632M & 51.941M \\

Accuracy
 & 86.62\% $\pm$ 2.8 & \textbf{92.93\% $\pm$ 2.7} & 88.78\% $\pm$ 3.2
 & 45.09\% $\pm$ 0.5 & 46.25\% $\pm$ 0.652 & \textbf{60.60\% $\pm$ 1.852} \\

\hline
\end{tabular}
\caption{
Artificial network performance on the context resolution task using the temporal GSC as a base dataset (abbreviated Seq GSC).
A varying number of gaps in the range $[0, 30]$ is introduced during training, and evaluation uses a fixed gap duration of $30$. 
Parameters for the sequences with a complexity of $\mathrm{TE} = 1.71$: $|\Sigma| = 15$, $14$ ambiguities with maximum depth $15$, transition density = $0.02$, $|q_0| = 4$,
$\mathrm{len}(S_i) \in [1, 30]$.
Parameters for the sequences with a complexity of $\mathrm{TE} = 2.61$: $|\Sigma| = 35$, $34$ ambiguities with maximum depth $35$, transition density = $0.01$, $|q_0| = 4$, $\mathrm{len}(S_i) \in [1, 60]$.
Results are reported as mean $\pm$ standard deviation over 4 runs.
}
\label{tab:ann_performance}
\end{table}

\subsection{Syntactic structure of animal behavior: analyzing latent dynamics from the perspective of generative linguistics}
\label{sec:behavior-syntax}

Beyond the generation of computational tasks and datasets of controllable complexity, \verb|SymSeqBench| provides a comprehensive set of analysis tools and metrics to quantify the complexity of symbolic sequences and infer the properties of the underlying generative processes. Given the scope and breadth of these metrics (see \prettyref{sec:metrics}), their application to empirical datasets can provide a valuable complement to understand the syntactic structure of different types of sequences across different scientific domains. In this section, we exemplify these applications and demonstrate the types of insights they can provide, using openly available datasets comprising behavioral sequences across different species and different behavioral categories.  

\begin{figure}
    \centering
    \includegraphics[width=\linewidth]{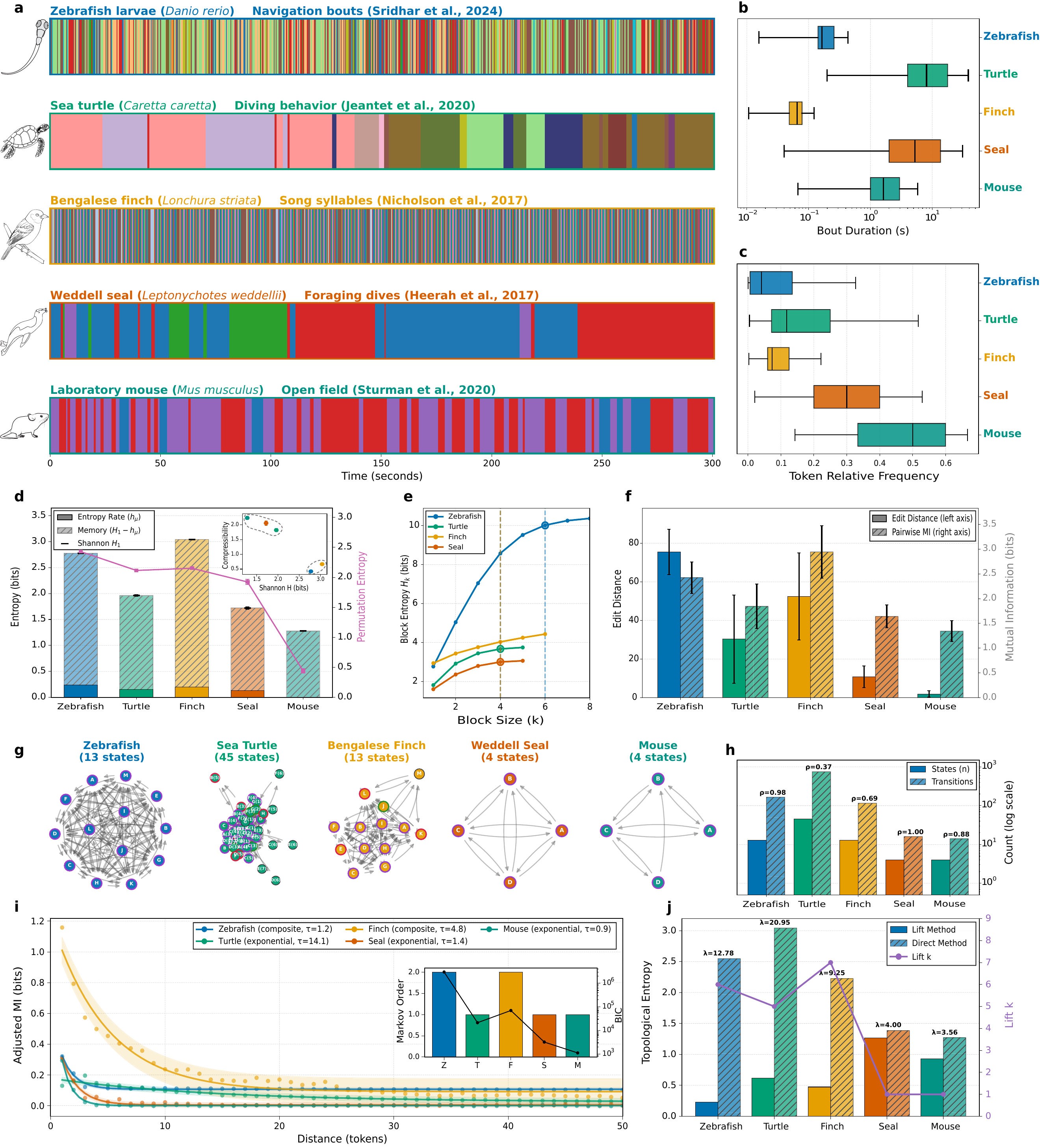}
    \caption{\textbf{Multi-scale analysis of behavioral sequence complexity across animal species.} Behavioral sequences from five species (zebrafish larvae, sea turtles, Bengalese finches, Weddell seals, laboratory mice) were analyzed at four hierarchical levels. \textbf{a)} Ethograms showing representative sequences with color-coded behavioral states over 300-second windows. \textbf{b-c) Token-level analysis:} bout duration distributions and relative frequencies reveal species-specific temporal patterns. \textbf{d-e) String-level analysis:} the entropy rate ($h_\mu$) captures stochastic complexity, while the memory component ($H_1 − h_{\mu}$) quantifies temporal dependencies, and Shannon entropy ($H_1$) represents total uncertainty (d, left axis). Permutation entropy (d, right axis) provides a complementary, model-free, complexity measure and string compressibility correlates strongly with Shannon entropy (d, inset). The growth of block entropy ($H_k$) as a function of block size (e) quantifies long-range dependencies with the marked saturation points revealing memory depth. \textbf{f) String-set-level} edit distance and pairwise mutual information quantify stereotypy. \textbf{g-j) Grammar-level analysis:} transition graphs (g) and their properties (h) can be estimated from first-order transition probabilities, while the existence and characteristics of non-adjacent and hierarchical relations can be inferred from information-based metrics (i), Markov order fitting quality (i, inset) or topological entropy measures (j). \textit{Data sources:} \cite{sridhar2024gradient} (zebrafish), \cite{jeantet2020combined} (turtle), \cite{nicholson2017bengalese} (finch), \cite{heerah2017validation} (seal), \cite{sturman2020deep} (mouse). \textit{Animal illustrations:} \cite{costa2021zebrafish} (zebrafish), \cite{tyler2019mouse} (mouse) from SciDraw under CC-BY 4.0; turtle, finch, and seal from Vecteezy.com.}
    \label{fig:animal-syntax}
\end{figure}

Bolstered by advanced machine learning tools that automate tracking and positional estimation from recorded behavioral observations (e.g. DeepLabCut \cite{mathisDeepLabCutMarkerlessPose2018, yeSuperAnimalPretrainedPose2024}, DeepOF \cite{mirandaDeepOFPythonPackage2023} or SLEAP \cite{pereiraFastAnimalPose2019}), combined with the streamlined identification, annotation and labeling of discrete behavioral bouts and sequences \cite{weinrebKeypointMoSeqParsingBehavior2024, linCharacterizingStructureMouse2024}, the emerging new scientific discipline of computational neuroethology \cite{Datta2019, vonzieglerBigBehaviorChallenges2021} can revolutionize how we understand naturalistic animal behavior and the characteristics of the underlying generative processes, providing a new, integrative view on brain, cognition and behavior. 

By focusing on abstract and generic symbolic sequences, while retaining close inspiration from psychology and cognitive sciences, the tools and metrics provided by \verb|SymSeqBench| can be a valuable complement to behavioral analysis pipelines (as well as other classes of empirical data that can be expressed as symbolic sequences). To illustrate the scope of analysis metrics and tools, we analyzed behavioral sequences from five different species (obtained from openly available datasets) at four hierarchical levels (see \prettyref{fig:animal-syntax} and the formal description of all metrics in \prettyref{sec:metrics}): token, string, string-set, and grammar. While token-level analysis primarily emphasize differences in distributional statistics and temporal properties of individual behaviors (with Zebrafish swimming and Finch's songs showing shorter and less frequent syllables, \prettyref{fig:animal-syntax} a-b), string- and string-set-level analyses reveal differences in contextual depth and complexity. Decomposing the entropy of individual strings into a stochastic component (entropy rate, $h_\mu$) and a memory component (residual entropy, $H_1 - h_\mu$) allows us to quantify both the total uncertainty / randomness in the observed behavioral sequences (\prettyref{fig:animal-syntax} d, left axis) and how much of that uncertainty is due to higher-order temporal dependencies (\prettyref{fig:animal-syntax} d, stacked bars). The permutation entropy then provides a complimentary, model-free complexity measure ((\prettyref{fig:animal-syntax} d, right axis) re-inforcing the overall pattern: mouse open field behavior is the simplest, seal and turtle show stereotyped and low-entropy behavioral sequences, whereas zebrafish and finch tend to exhibit higher complexity. These metrics also align with sequence compressibility (\prettyref{fig:animal-syntax} d, inset), yielding a clear grouping between high-entropy, low-compressibility (complex) sequences for zebrafish and finch versus low-entropy, high-compressibility (simple) sequences for the other datasets. If we assess how the sequence entropy changes if we chunk the sequences into increasingly larger groups (\prettyref{fig:animal-syntax} e), the same pattern is evident with zebrafish behaviors showing longer contextual dependencies (saturating at $k=6$), turtles and seals showing the shortest ($k=4$). It is worth noting that this analysis also emphasizes limitations of these metrics that would require longer individual strings to yield meaningful conclusions for the two remaining datasets. The complementary string-set complexity metrics (\prettyref{fig:animal-syntax} f) then quantify how stereotypical different behavioral epochs / strings are compared to the others in the string set, demonstrating the same pattern: simple, stereotypical behaviors for mouse, seal and turtle; complex, flexible behaviors for zebrafish and finch. 

Beyond statistical quantification and structural analyses, the set of metrics we implement allow us to identify contextual dependencies and infer the properties of the generative grammars underlying observed behaviors. First-order transition probability graphs and corresponding properties (state and transition densities, \prettyref{fig:animal-syntax} g-h) vary according to the complexity of the observed behavioral repertoire but prove insufficient to capture the full complexity of generative grammars (\prettyref{fig:animal-syntax} i, j). These analyses directly confront a fundamental challenge in formal language theory: the \textit{grammatical inference problem}, which asks whether we can unambiguously determine the generative grammar---and specifically, its position within the Chomsky hierarchy (regular, context-free, context-sensitive, or recursively enumerable)---from observed sequences alone. Seminal theoretical work has established fundamental limitations on this endeavor. Gold's theorem \cite{goldLanguageIdentificationLimit1967} demonstrated that no superfinite class of formal languages is identifiable in the limit from positive examples alone (i.e., without access to negative examples indicating which strings are \textit{not} part of the language). This negative result extends even to the class of regular languages, the simplest level of the Chomsky hierarchy \cite{angluinInductiveInferenceFormal1980}. Furthermore, even when the target grammar class is known a priori and restricted to regular languages, finding the minimal (smallest) consistent automaton or grammar is NP-hard \cite{pittMinimumConsistentDFA1989}, and the approximation within constant factors remains computationally intractable \cite{charikarSmallestGrammarProblem2005,lehmanApproximationAlgorithmsGrammarBased2002}. For context-free and more expressive grammars, complexity increases dramatically: exact inference is undecidable in the general case, and even for learnable subclasses such as substitutable context-free languages \cite{clarkPolynomialIdentificationLimit2007}, polynomial-time algorithms require strong structural constraints and access to characteristic samples that may not be available in naturalistic behavioral data.

These theoretical barriers force us to adopt a pragmatic approach to analyzing sequences: rather than attempting to extract a precise minimal grammar (which is computationally infeasible), we employ a battery of complementary statistical metrics to \textit{characterize} sequence complexity, long-range dependencies, and structural properties that collectively provide evidence for or against membership in specific grammar classes. This methodology is flawed, trading rigour for feasibility, but, by employing multiple complementary and independent metrics it can provide convergent evidence to make informed inferences about the underlying generative processes.

Sequential, adjusted mutual information (MI) decay (adapted from \citep{Lin2017,Sainburg2019,Sainburg2022} \prettyref{fig:animal-syntax} i) quantifies long-range statistical dependencies; rapid exponential decay in seals and mice indicates short-term correlations characteristic of low-order Markov processes, while extended, composite decay in finches and zebrafish reveals longer-range, hierarchical structure consistent with supra-regular (beyond finite-state) grammars. This interpretation is validated by Markov order estimation (inset), which demonstrates that sequences observed for turtles, seals, and mice can be adequately modeled as first-order Markov processes (low BIC scores), whereas zebrafish and finch datasets require longer contextual dependencies and are not well approximated by low-order Markov models, suggesting non-Markovianity and potentially hierarchical structure. These metrics are then complemented by topological entropy (TE) analysis (\prettyref{fig:animal-syntax} j) which provides a canonical, independent, model-free measure of generative capacity \citep{Robinson1998,Bollt2000}. Seal and mouse data exhibit convergence with the lift method at depth 1 (consistent with first-order Markov processes and regular grammars), whereas zebrafish, turtle, and finch datasets require larger contextual windows and show systematic discrepancies between direct computation and lift-based approximation. These divergences suggest that simple finite-state models cannot adequately capture the generative complexity, pointing toward context-free or higher-order dependencies. Importantly, the convergence values ($\lambda$) provide a quantitative complexity ordering: zebrafish exhibit the highest TE ($\lambda = 20.95$), reflecting rich navigational repertoires with extensive state spaces, while seals and mice show the lowest ($\lambda = 4.00$ and  $3.56$), consistent with stereotyped, low-entropy foraging routines.

Collectively, these grammar-level analyses do not definitively resolve the class membership of observed behavioral sequences within the Chomsky hierarchy, an objective that formal language theory has proven to be fundamentally unattainable without additional constraints. However, the convergent patterns across multiple independent metrics (MI decay timescales, Markov order estimates, topological entropy convergence properties, and transition graph structures) provide principled, probabilistic evidence that zebrafish navigation and finch song likely involve supra-regular generative processes (consistent with what was proposed in the corresponding original publications), while seal, turtle, and mouse behaviors appear consistent with regular or low-order Markovian grammars. This inferential framework, grounded in rigorous computational constraints, represents a pragmatic and methodologically sound approach to characterizing the syntactic complexity of naturalistic behavioral sequences.

\section{Discussion and Outlook}
\label{sec:discussion}
Sequences act as the shared currency across cognitive processes, behavioral repertoires, and modern AI architectures, offering unified means to examine the structural dependencies that shape natural and artificial information processing. Leveraging this common structure demands benchmarks that are cognitively inspired, theoretically principled and methodologically flexible, capable of probing sequence complexity in a controlled yet comparable way. However, current resources remain fragmented and lack a shared representational foundation: large-scale language benchmarks are costly and opaque with respect to the specific abilities they test, while lightweight synthetic tasks rarely capture the spatio-temporal diversity and compositional structure central to human and animal cognition. \verb|SymSeqBench| aims to overcome these challenges through an open-source initiative that offers a principled framework for generating, transforming, and analyzing symbolic sequences with controllable complexity. By unifying dataset construction, task specification, and multi-scale complexity analysis through low-barrier and user-friendly interfaces, we expect the tool to engage researchers from a variety of disciplines interested in systematic, cross-domain analysis and benchmarking of cognitive and computational models.

\verb|SymSeqBench| taps into computability and learnability theories and uses the less-known but powerful TE metric to efficiently sample user-constrained formal grammars of desired complexity, filling a key gap in synthetic data generation tools and providing new benchmarks for biologically-detailed, artificial and neuromorphic systems. Complementing but not replacing full-fledged parsers \citep{bird2009natural,Silberztein2016} and grammar induction libraries \citep{solan_unsupervised_2005,nawrocki_infernal_2013} that typically deal with large text corpora or genome data, for example, it focuses on simple and interpretable structures and allows their manipulation in behaviorally and computationally relevant ways (e.g., by introducing deviants). Together with many pre-built cognitively inspired paradigms (e.g., NAD learning or AGL) and flexible symbolic- and embedding-level task specifications, \verb|SymSeqBench| enables methodical investigation of statistical and rule-based learning with direct links to a rich experimental, behavioral and theoretical literature. Conversely, linguists and cognitive scientists can benefit from the automatic generation of experiment-ready, balanced AGL datasets from arbitrary grammars at an unprecedented level of fine-grained control and scale. To the best of our knowledge, no other actively maintained software offers comparable functionality: AGL Stimselect is more limited and unsupported, and AGSuite provides only a web-based feature-analysis interface.

We nevertheless expect \verb|SymSeqbench|'s primary use-case to be the evaluation of neural network models across architectures. To this end, it includes efficient routines for mapping symbols onto either synthetically generated embeddings or established datasets while exposing flexible transformations and property-manipulations during runtime. Our tool thus addresses a critical gap in evaluating sequential processing capabilities: while large language models are typically assessed on complex, naturalistic benchmarks \citep{srivastava2023imitationgamequantifyingextrapolating, wang2020supergluestickierbenchmarkgeneralpurpose,liang2023holisticevaluationlanguagemodels}, these evaluations make it difficult to isolate specific computational abilities or systematically control task complexity \citep{Raji2021,ribeiro2020accuracybehavioraltestingnlp}. Conversely, targeted synthetic tasks designed for recurrent neural networks \citep{Graves14_ntm,weston2015memorynetworks} or biologically-detailed spiking neural networks \citep{Eshraghian2023,SHD} often lack the scope and standardization needed for cross-architecture comparison. \verb|SymSeqBench| bridges this divide by enabling controlled evaluation across the full spectrum of neural architectures -- from feedforward networks and recurrent architectures (LSTMs, GRUs) \citep{Cho14_GRU} to attention-based transformers and large language models \citep{Deletang22_chomsky,Vaswani17_attention} -- using tasks with systematic complexity control, explicit theoretical grounding and clear relations to Human cognitive performance.

This positions our tool uniquely between libraries focusing on dataset preprocessing \citep[e.g., tonic][]{Lenz21_5079802}, reinforcement-learning (RL) environments like NeuroGym \citep{manuel_molano_maz_n_c01c84dc}, full-stack SNN frameworks (Norse \citep{Pehle_Norse_-_A_2021}, SpikingJelly \citep{Fang2023}, SNNTorch \citep{Eshraghian2023}, Jaxley \citep{Deistler2025}) or neural simulators (NEST \citep{Gewaltig:NEST}, Brian2 \citep{Stimberg2019}), or applications designed for human experiments (PsychoPy \citep{Peirce2019}). Although \verb|SymSeqBench| shares certain capabilities with these tools, it remains distinct in scope and design by providing a unified framework for systematic sequence complexity control applicable across all neural modeling paradigms.

Whenever possible, it leverages and builds on high-quality software to benefit from advances in the deep learning community, e.g., for dataset handling (tonic \citep{Lenz21_5079802}) and transformations (torchvision \citep{TorchVision} and torchaudio \citep{hwang2023torchaudio}).
Beyond its central emphasis on temporal processing, the framework smoothly integrates and transitions between discrete vector embeddings, continuous-time signals and spiking encodings, enabling efficient integration with Torch-based models while also providing generic output formats compatible with any tool.
Unlike NeuroGym or PsychRNN \citep{ehrlich_psychrnn_2021}, which mostly focus on decision-making or delayed-response paradigms with predefined input structure and little control over task difficulty, \verb|SymSeqBench|'s symbolic framework allows defining more general sequence-processing tasks while retaining maximum flexibility on the input embeddings. In contrast to some approaches (conn2res \citep{suarez_connectome-based_2024}, PsychRNN), it deliberately avoids tight integration with specific simulation engines, which leads to some user overhead but also guarantees complete freedom of backend choice and shields the tool from premature obsolescence \citep[e.g., Theano-based PyCog][]{song_training_2016}. 

Decoupling the task formulation from the symbol embeddings is a novel approach that has important conceptual and practical implications: it allows the same task to be evaluated using embeddings/datasets (explicitly specified by users) of various complexities, which is known to impact generalization capacity \citep{Poggio2018,kamkari_geometric_2024}; provides direct support for multi- and cross-modal experiments, a key line of research for studying multisensory integration and cross-modal generalization underlying abstract representation learning in biological \citep{okray_multisensory_2023,senkowski_multi-timescale_2024,guyoton_cortical_2025} and artificial  \citep{ngiam2011multimodal,ramachandram_deep_2017,ito_generalization_2024} systems; and allows \verb|SymSeq| and \verb|SeqBench| to be used separately, e.g., for generating only AGL datasets versus attributing a meaning to the symbols and evaluating computational models. 
The modular architecture naturally blends high customizability with fast prototyping, allowing the  deployment of complete data generation pipelines using simple configuration files (suitable for non-coders), thereby letting researchers focus more on the models, shortening experimental round-time and ensuring fully reproducible \citep{pauli_reproducing_2018, benureau_re-run_2017} datasets.

Beyond synthetic task generation, \verb|SymSeqBench| provides a comprehensive suite of analysis tools applicable to both generated and empirical sequences. The multi-scale metric framework -- spanning token, string, string-set, and grammar levels -- enables the systematic characterization of sequence complexity across diverse domains, from animal behavioral to neural activity patterns and linguistic corpora. Any source of empirical data that can be cast as a set of symbolic sequences is amenable to analysis within this framework.

At the grammar level, topological entropy (TE) serves as the primary structural complexity measure, quantifying the asymptotic exponential growth rate of unique grammatical strings as a function of length \citep{Robinson1998,Bollt2000}. Unlike compression-based metrics (Lempel-Ziv, Kolmogorov complexity) that emphasize statistical regularities or string entropy metrics that emphasize local variability, TE captures the combinatorial richness of the grammar by analyzing its adjacency structure providing a fast, precise, and theoretically grounded measure that can be used both for generation (creating regular grammars of controlled complexity) and analysis (determining the complexity of generative rules). Our comparative analyses demonstrate that TE exhibits superior sensitivity to structural variations (e.g., ambiguity, clustering, transition density) compared to compression ratios or perplexity-based measures, making it especially suited for systematic control and manipulation of dataset complexity. 

However, TE has inherent limitations that must be acknowledged. First, it applies exclusively to \textit{regular grammars} (finite-state automata); context-free and more expressive grammars require alternative complexity measures or approximations (e.g., probabilistic variants or upper bounds). Second, TE quantifies only the \textit{potential} for generating diverse strings and does not account for statistical biases in actual string distributions -- two grammars with identical TE may produce vastly different empirical sequence statistics if certain paths are favoured. Third, the direct eigenvalue method assumes access to the complete transition structure; for inferred or partially observed grammars, estimation errors propagate into TE calculations, potentially yielding misleading complexity assessments.

The string-level and corpus-level metrics complement TE by capturing statistical, distributional, and information-theoretic properties that reflect how complexity manifests in observed sequences. Entropy decomposition (entropy rate, memory component, Shannon entropy) quantifies stochastic complexity and temporal dependencies; edit distance and mutual information assess inter-sequence similarity and stereotypy; perplexity and divergence measures evaluate model fit and distributional shifts. This multi-metric approach provides convergent evidence for characterizing sequence structure, mitigating the limitations of any single measure.

Crucially, these analysis tools are entirely decoupled from the generation pipeline, enabling their application to arbitrary symbolic time-series data. Users can import empirical sequences - from animal behavioral ethograms, neural population activity latent states, or time-series data converted to symbolic form via SAX (Symbolic Aggregate approXimation) \citep{lin_symbolic_2003} - and apply the full metric suite to quantify complexity, infer approximate grammar classes, or compare statistical properties across conditions. This positions \verb|SymSeqBench| as a versatile analytical toolkit for computational neuroethology, cognitive neuroscience, machine learning and artificial intelligence, as well as any other domain where sequential structure plays a central role.

\subsection{Grammar inference and the decidability constraint}

A natural application domain of \verb|SymSeqBench|'s analysis capabilities is the inference of generative grammars from observed sequences. This task has profound implications for understanding the computational principles underlying sequential structure, inferring the characteristics of generative mechanisms and deploying rigorous computational benchmarks. However, the objective confronts fundamental theoretical limits established by formal language theory and computational learning theory. Gold's theorem \citep{goldLanguageIdentificationLimit1967} proved that no superfinite class of formal languages is identifiable in the limit from positive examples alone, even for regular languages \citep{angluinInductiveInferenceFormal1980}. Moreover, finding the minimal consistent automaton for a given set of strings is NP-hard \citep{pittMinimumConsistentDFA1989}, and approximation within constant factors remains intractable \citep{charikarSmallestGrammarProblem2005,lehmanApproximationAlgorithmsGrammarBased2002}. For context-free and more expressive grammars, exact inference is undecidable in the general case, and even learnable sub-classes require strong structural constraints (e.g., substitutability) and characteristic samples that may not be available in empirical data \citep{clarkPolynomialIdentificationLimit2007,delaHigueraCharacteristicSetsPolynomial1997}.

This impossibility does not invalidate grammar inference as a research goal; rather, it clarifies the boundaries of what is computationally achievable and motivate pragmatic methodologies. Instead of attempting to extract the true minimal grammar, \verb|SymSeqBench|'s multi-scale analysis framework provides \textit{probabilistic evidence} for grammar class membership and structural properties through consensus patterns across independent metrics. For example, rapid mutual information decay, low Markov order, and TE convergence at depth 1 collectively suggest a regular (finite-state) grammar, while extended MI decay, high block entropy growth, and TE divergence between direct and lift methods indicate supra-regular structure (context-free or beyond). This approach trades exactness for feasibility, leveraging computational tractability to make informed, evidence-based inferences about generative processes.

Currently, \verb|SymSeqBench| provides basic grammar inference capabilities for regular languages through automaton induction from positive examples, suitable for exploratory analysis and hypothesis generation. Future development will incorporate more sophisticated inference algorithms (e.g., state-merging methods, probabilistic grammar induction, Bayesian approaches) and extend support to restricted context-free sub-classes, enabling richer structural modeling while respecting computational constraints. Users should interpret inferred grammars as \textit{hypotheses} rather than ground truth, validated through cross-metric consistency checks and generalization performance on held-out data.

\subsection{Future directions}

While \verb|SymSeqBench| already provides a comprehensive foundation for symbolic sequence generation and analysis and a solid set of controllable tasks to evaluate and compare cognitively-inspired computing, there are several extensions under active development. Although some tasks already employ hierarchical and recursive structures, central to natural language syntax, our ability to analyze, induce and infer supra-regular grammars is currently limited in scope. Pushing towards that direction will enable the development of more complex compositional reasoning tasks and, consequently, the evaluation of more naturalistic cognitive computations. This includes both CFG-based sequence generation with controllable complexity (via probabilistic CFGs and entropy-based constraints) and inference of CFG sub-classes from empirical data, building on recent advances in distributional learning \citep{clarkPolynomialIdentificationLimit2007} and Bayesian grammar induction. 

Regarding the symbolic embeddings implemented in \verb|SeqBench|, there is ample space for integration with other established datasets such as TIMIT (phoneme recognition), other spoken digit corpora, and other psycholinguistic paradigms (e.g., overlap, cohort, and TISK tasks probing lexical access and temporal integration). The software is modular and developed in a highly extensible manner that will facilitate the expansion and integration of these and future datasets. Additionally, current embedding decoupling already enables cross-modal experiments (e.g., visual-auditory sequence learning), but future versions will provide tighter integration with multimodal datasets and transformations, explicit support for alignment/synchronization of multi-stream sequences, and tools for quantifying cross-modal statistical dependencies and generalization.

Regarding metrics, ongoing work includes developing robust TE approximation methods for non-regular grammars (e.g., via probabilistic bounds or hierarchical decomposition), uncertainty quantification for TE estimates from inferred grammars, and alternative complexity measures tailored to specific grammar classes (e.g., nested depth for Dyck languages, crossing dependencies for mildly context-sensitive grammars). We also aim to expand upon the categorization of generative grammars into the Chomsky hierarchy, providing a detailed scheme to determine the likelihood of an observed (or generated) sequence being caused by grammars at different complexity levels. 

To improve accessibility for non-programmers and facilitate exploratory analysis, we are developing web-based interfaces for grammar visualization (state transition diagrams, derivation trees), real-time metric computation and comparative visualization, and interactive parameter tuning for grammar generation and task configuration. As an open-source project, \verb|SymSeqBench| thrives on community contributions. We actively encourage users to share custom grammars, task definitions, analysis metrics, and empirical datasets through a centralized repository, fostering reproducibility and accelerating methodological innovation across disciplines. Ongoing feedback from cognitive scientists, neuroscientists, and machine learning researchers will guide prioritization of feature development and ensure the tool remains responsive to evolving research needs.

In summary, \verb|SymSeqBench| provides a set of standardized, cognitively grounded benchmarks for evaluating temporal processing in both biological and artificial neural networks, the tools to generate stimulus sets for experimental psychology, as well as metrics to evaluate structural complexity of any sequentially structured dataset. Bridging the gap between highly controlled synthetic tasks, naturalistic language processing and rigorous defining characteristics of human linguistic capacities, this tool aims to approximate artificial and natural intelligence.
By addressing these issues while maintaining modularity and ease of use, \verb|SymSeqBench| aims to establish itself as a standard resource for sequence processing research, bridging theoretical linguistics, cognitive neuroscience, and artificial intelligence through a shared computational and experimental framework. 

\section{Methods}

\subsection{Network architectures}

\paragraph{Biological neural networks.}
We consider two biologically-plausible spiking network models of sequence learning proposed by \citet{cone_learning_2021} and \citet{asabuki_neural_2022}, abbreviated as CS and AKF, respectively. Since the models are described in full detail in the original publications, here we only provide a brief description and focus more on modified parameters and task evaluation.

The CS model relies on a columnar architecture composed of Timer and Messenger cells (modelled as LIF neurons) to learn both transitions between sequence items and the duration of the individual elements, using a reward-modulated learning rule. The spiking version of the CS model was designed for deterministic sequences obeying the Markov property, while handling contextual dependencies requires memory-augmented architectures, which were demonstrated in continuous-rate networks but which are difficult to achieve in spiking networks. Here we used the NEST-based implementation \citep{NEST3.4} from \citet{Zajzon2023} -- which corrects an error in the original implementation --, retaining all parameter values with the exception of the inter-stimulus interval (ISI), which was set to $250$ ms in all tasks. 

The AKF model consists of a fully connected, non-Dalean (i.e., without separate excitatory and inhibitory units) network of stochastic spiking neurons, each composed of a dendritic and a somatic compartment. Through a combination of recurrent gating and unsupervised synaptic plasticity, this model can learn context-dependent segmentation of spike pattern sequences. For the implementation, we relied on the Python source code provided by the authors and kept all parameter values.

Model evaluation is based on the reservoir computing approach, sampling the population responses at the stimulus offset (not including the ISI) and training linear readouts with ridge regression on task-specific, one-hot encoded target labels (tokens). The state variable for both models is the instantaneous firing rate, and for CS we used only the rates of the Timer cells across all columns. As the imbalance in the token frequency can bias the accuracy and make it difficult to estimate a random baseline, we instead opted for the Cohen's kappa statistic \citep{Cohen1960} as a performance measure. This metric takes values between −1 and 1, 0 being chance level and values towards 1 representing good performance.

In the classification tasks in \prettyref{fig:app-bio-nn}a,b, the input strings are composed of the alphabet's (non-indexed) symbols (see main text for parameter values). For the n-step memory task, the target labels correspond to the identity of n-th preceding (non-indexed) symbol,  with valid targets restricted to lie within each individual string (i.e., no targets referring to previous strings). In contrast, the context resolution task uses the indexed states as target labels. The effective training and testing data sizes are model-specific and chosen to yield the best possible performance. Specifically, each epoch consists of a set of randomly drawn grammatical strings such that the total number of tokens (approximately) corresponds to the model size $M$ (number of neurons used for readout training), with $M_\mathrm{CS} = 500$ and $M_\mathrm{AKF} = 1200$. This ensures that one epoch contains suﬀicient data points to avoid overfitting of the readout. We set the total number of training epochs $\mathcal{E}_\mathrm{train}$ = 10 and the test size to $\mathcal{E}_\mathrm{test} = 0.2\mathcal{E}_\mathrm{train}$.

For the counting task in \prettyref{fig:app-bio-nn}c, the indexed symbols (e.g., $\mathrm{A}_\mathrm{1}$ or $\mathrm{X}_\mathrm{3}$) denote distinct stimuli. For each string, there is a single target defined for the last filler element $\mathrm{X}_\mathrm{j}$ and corresponding to the correct dependent item $\mathrm{B}_\mathrm{i}$. Here an epoch consists of single presentations of all unique strings allowed by the task parameters, and we set $\mathcal{E}^\mathrm{CS}_\mathrm{train}=250$ and $\mathcal{E}^\mathrm{AKF}_\mathrm{train}=500$, as well as $\mathcal{E}_\mathrm{test} = 100$ for both models.

\paragraph{Artificial and neuromorphic neural networks.}

We first consider two standard spiking neuron models for sequence processing, the leaky integrate-and-fire (LIF) model and its adaptive extension (adLIF), as defined in \citep{Bittar22}. At timestep $t$, $I_t$ denotes the input signal, $s_t$ the binary spike, and $\theta$ the firing threshold.
The discretized membrane potential $U_t$ of the LIF neuron evolves as:
\begin{align}
    U_t &= \alpha (U_{t-1} - \theta s_t) + (1 - \alpha) I_t, \\
    s_t &= \mathbf{1}(U_t \ge \theta),
\end{align}
where $\alpha < 1$ is the decay rate.
The adLIF model includes an intrinsic adaptation variable $w_t$ that influences the membrane potential by capturing recent neuronal activity, incorporating both spike-triggered and sub-threshold adaptation. 
Its discrete dynamics are given by:
\begin{align}
    U_t &= \alpha (U_{t-1} - \theta s_{t-1}) + (1 - \alpha)(I_t - w_{t-1}), \\
    w_t &= \beta w_{t-1} + a\, U_{t-1} + b\, s_{t-1}, \\
    s_t &= \mathbf{1}(U_t \ge \theta),
\end{align}
where $\alpha$ and $\beta$ are rate constants. The parameters $\beta$, $\alpha$, $a$, and $b$ are learnable and optimized during training. 
Neurons are stacked in $N_\text{layers}$ feed-forward layers, each comprising $N_{\text{hidden}}$ neurons. An output layer of leaky integrators with a learnable decay rate $\alpha$ produces the final outputs.
In all experiments, we perform a grid search over the following hyperparameters: 
$N_{\text{layers}} \in \{2,3,4,5\}$, 
$N_\text{hidden} \in \{128,256,512\}$, and
$\theta \in [0.0,1.5]$ with step 0.01.
Based on the hyperparameter search, the largest number of layers consistently performed best; therefore, all reported experiments use an architecture with 5 layers and a hidden size of 256.
\par
In addition, we assess three representative sequence processing artificial neural networks on more challenging sequences: gated recurrent units (GRU) \citep{Cho14_GRU}, the transformer decoder \citep{Vaswani17_attention}, and the selective state space model Mamba \citep{Gu23_mamba}.
GRUs process sequences by recurrently updating a hidden state as each new token is processed, utilizing input and state-dependent gates to control how much past information is retained or overwritten. Despite their efficiency and performance on small sequences, GRUs have shown limited capacity for learning long contexts and are relatively slow to train.
The attention mechanism, at the core of the Transformer \citep{Vaswani17_attention}, improves long-sequence processing by replacing recurrence with a global memory of past token representations, which are retrieved through query, key, and value interactions. This enables direct modeling of long-range dependencies, but incurs quadratic time and memory complexity in sequence length.
State space models such as Mamba \citep{Gu23_mamba} revisit recurrent modeling through a state space formulation that employs linear recurrences for efficient training. Careful initialization and state expansion increase effective memory capacity, resulting in strong long-sequence performance with linear-time complexity.
The three different architectures are configured as follows: the Mamba model utilizes a state space expansion factor of 64, a local convolution width of 4, and a block expansion factor of 2. The GRU architecture is configured as a single-layer unidirectional recurrent network with hidden size matching the model dimension. The Transformer decoder employs 4 attention heads with rotary positional embeddings (RoPE base frequency of 10000), RMSNorm for layer normalization, and a SwiGLU feedforward network. All models use a dropout rate of 0.2 and LayerNorm for normalization between layers. The input sequences are first projected to the model through a linear encoder, and the final outputs are produced via a linear decoder layer.
\par
Both the ANNs and neuromorphic models are benchmarked on classifying the indexed states underlying the input sequences (which corresponds to a next-token prediction task for ambiguous sequences, see main text for more details). For the ANN experiments, we use 200000 training samples and 10000 test samples, and train the models for 50 epochs. For the SNN experiments, we also use 200000 training samples but only 1000 test samples, and train for a single epoch due to the computational overhead of SNN training. During training, a cross-entropy loss is computed at every time step of each sample, including the gap duration, by predicting the indexed state associated with the current sequence position. Gradients are propagated through the full sequence using backpropagation through time. Model performance is evaluated by measuring classification accuracy exclusively during the gap period.
For the optimization of the ANNs, we used the Adam optimizer, with an initial learning rate of 0.001, and the learning rate following a cosine annealing schedule with a warmup period corresponding to 5\% of the total training steps.
For the SNN experiments, we also used the Adam optimizer, with a fixed learning rate of 0.005 and no learning rate scheduler.

\subsection{Metrics}
\label{sec:metrics}

To enable a comprehensive characterization of sequential data, either internally generated or externally acquired, \verb+SymSeqBench+ includes a comprehensive, hierarchical suite of complexity metrics organized across four levels of structural granularity: token-level, string-level, string-set level, and grammar-level metrics. This taxonomy provides a systematic framework for quantifying different aspects of sequence complexity, from local statistical properties to global generative capacity.

\paragraph{Token-level metrics} characterize the statistical and distributional properties of individual symbols within sequences. Basic measures include \textbf{token frequency} ($f(\sigma_i, S) = \#\{\sigma_i \in S\} / |S|$), which quantifies symbol distribution uniformity, and \textbf{token duration statistics}, capturing temporal persistence through mean duration and variance across symbol occurrences. For sequences with embedded token representations, we compute \textbf{embedding complexity} metrics including vector norms, spectral radius of the embedding matrix, and pairwise embedding distances, which characterize the geometric structure of the symbolic input space.

\paragraph{String-level metrics} quantify complexity properties of individual sequences or concatenated string-sets, capturing information content, compressibility, and temporal dependencies. \textbf{Shannon entropy} ($H(S) = -\sum P(\sigma_i) \log_2 P(\sigma_i)$) provides a foundational measure of uncertainty in symbol distributions \citep{Shannon1948}. We extend this by decomposing entropy into \textbf{block entropy} ($H_L(S)$), computed over $L$-grams, and the asymptotic \textbf{entropy rate} ($h_\mu(S) \approx H_L(S) - H_{L-1}(S)$), which captures the information content per symbol for long sequences. The \textbf{Effective Measure Complexity} (EMC; \citep{Grassberger1986}) quantifies deviation from maximal randomness through $\text{EMC}(S) = \sum [H_L(S) - L \times h_\mu(S)]$, measuring structural regularity beyond simple entropy.

Algorithmic complexity is assessed through compression-based metrics. \textbf{Compressibility} using standard deflate/gzip compression ($C_{\text{gzip}}(S) = |C(S)| / |S|$) and \textbf{Lempel-Ziv-Welch complexity} \citep{Lempel1976,Welch1984} quantify sequence redundancy and repetition structure. \textbf{Linguistic complexity} captures vocabulary richness through the product of normalized unique $n$-gram counts across scales: $C_{\text{ling}}(S) = \prod U_i(S)$, where $U_i(S) = |V_i(S)| / \min(|A|^i, |S|-i+1)$ \citep{Trifonov1990}. \textbf{Permutation entropy} \citep{Bandt2002} further characterizes ordinal pattern distributions, providing a non-parametric measure of temporal structure invariant to monotonic transformations.

\paragraph{String-set level metrics} characterize relationships and variability across collections of sequences. Pairwise sequence distances include the \textbf{Hamming distance} ($d_H(S_i, S_j) = \sum \mathbbm{1}[S_i[k] \neq S_j[k]]$) for equal-length strings and \textbf{edit distance} (Levenshtein distance), computed via dynamic programming as the minimum number of insertions, deletions, and substitutions required for transformation. The \textbf{Normalized Compression Distance} (NCD; \citep{Cilibrasi2005}) provides a similarity metric based on algorithmic information theory: $\text{NCD}(S_1, S_2) = (C(S_1 \circ S_2) - \min(C(S_1), C(S_2))) / \max(C(S_1), C(S_2))$. More specifically tailored to psycholinguistic studies, \textbf{Associative chunk strength} (ACS; \citep{Knowlton1996,Bailey2008}) quantifies the familiarity of $n$-gram subsequences based on their frequency in training data, with global and anchor-based variants capturing different aspects of sequential predictability. \textbf{Mutual information} between string pairs ($I(S_i; S_j) = H(S_i) + H(S_j) - H(S_i, S_j)$) and \textbf{string-set entropy} ($H(\mathcal{S}) = -\sum P(S) \log_2 P(S)$) characterize information-theoretic dependencies and diversity at the string-set level, with conditional variants capturing prefix-based structure.

\paragraph{Grammar-level metrics} characterize the underlying generative mechanisms and computational capacity of sequence-producing systems. \textbf{Markov order estimation} employs information-theoretic model selection criteria (Akaike Information Criterion, Bayesian Information Criterion) to determine optimal dependency range, assuming Markovianity: $\hat{k} = \argmin_k \text{BIC}_k$, where $\text{BIC}_k = -2 \log L_k + p_k \log N$ \citep{Csiszar2000,Katz1981}. For sequences exhibiting variable-length dependencies, we employ \textbf{Variable-Length Markov Models} (VLMC; \citep{Buhlmann1999}) to adaptively prune context trees using statistical significance criteria.

\textbf{Hierarchical dependency structure} is revealed through mutual information decay profiles across element distances, distinguishing exponential ($\text{MI}(d) \sim a \cdot e^{-bd}$) from power-law ($\text{MI}(d) \sim a \cdot d^{-b}$) decay signatures indicative of different organizational principles \citep{Lin2017,Sainburg2019,Sainburg2022}.

\textbf{Topological entropy} (TE) quantifies the exponential growth rate of distinct sequences as a function of length, providing a canonical measure of generative capacity \citep{Robinson1998,Bollt2000}. For a grammar $\mathcal{G}$ with $N$ unique states, let $M$ denote the $N \times N$ transition matrix satisfying the Markov property. TE is given by $h(G) = \log_e(\lambda_1)$, where $\lambda_1$ is the spectral radius (largest real eigenvalue) of $M$. This \textit{direct} eigenvalue approach enables fast, precise computation and relies on the indexed state representation to resolve symbol ambiguities \citep{Warren2015}. Alternative \textit{lifting techniques} \citep{Bollt2000} construct $N^l \times N^l$ matrices from $l$-grams but prove computationally impractical for complex grammars \citep{Warren2015}. Accurate TE calculation requires the transition matrix to be irreducible (strongly connected), aperiodic, and free of absorbing states - conditions necessary for eigenvalue-based methods to yield global, non-trivial entropy values \citep{Warren2015}. While typical artificial grammar specifications naturally satisfy irreducibility and aperiodicity, \verb|SymSeq| ensures the absence of absorbing states by adding a transition loop from the special end-of-sequence terminal symbol (\texttt{\#}) to all initial states, guaranteeing indefinite generation cycles. By the Perron-Frobenius theorem, the spectral radius is bounded from above by the maximum outdegree: $\text{TE}_{\max}(A_G) \leq \log(\max_{1 \leq i \leq n} \sum |A_{G_{ij}}|)$.

Additional grammar complexity measures include \textbf{production rule complexity} ($C_{\text{rules}}(G) = |R| + \sum |r|$, normalized by alphabet and non-terminal vocabulary sizes), \textbf{state complexity} (minimal deterministic finite automaton state count), and derivation tree statistics capturing average structural depth and branching \citep{Hopcroft2006}. For probabilistic grammars, \textbf{language entropy} ($H(L) = \lim_{n \to \infty} (1/n) H_n$, where $H_n = -\sum P(S) \log_2 P(S)$ over length-$n$ strings) quantifies uncertainty in the generative distribution. \textbf{Minimum Description Length} (MDL; \citep{Rissanen1978,Barron1998}) balances grammar compactness against data encoding efficiency: $\text{MDL}(G) = L(G) + L(D|G)$, providing a principled criterion for grammar induction and model selection.

This comprehensive metric suite enables fine-grained characterization of sequence complexity across multiple organizational scales, supporting rigorous comparison of datasets and systematic investigation of structure-learning relationships in cognitive and computational models.

\subsection*{Acknowledgements}
The authors gratefully acknowledge the computing time granted by the JARA-HPC Vergabegremium on the supercomputer JURECA at Forschungszentrum Jülich.

\subsection*{Funding}

This work was sponsored by the Federal Ministry of Education, Germany (project NEUROTEC-II grant no. 16ME0398K and 16ME0399), by Neurosys as part of the initiative "Cluster4Future", funded by the Federal Ministery of Education and Research BMBF (03ZU1106CB and 03ZU2106CB), by the Federal Ministry of Education and Research (BMBF) under grant no. 01IS22094E WEST-AI, by the Portuguese Foundation for Science and Technology (FCT) under projects UIDB/04539/2020, UIDP/04539/2020 and LA/P/0058/2020 and by the European Union’s Horizon 2020 research and innovation programme under grant agreement no. 952422 (DYNABrain).

\bibliographystyle{plainnat}
\bibliography{bib/refs_merged}

\end{document}